\documentclass[sn-mathphys, Numbered]{sn-jnl}
\pdfoutput=1
\usepackage{graphicx,caption,amsmath,amssymb}  
\usepackage{color,latexsym,array,multirow,verbatim,enumerate}

\usepackage{graphicx}%
\usepackage{multirow}%
\usepackage{amsmath,amssymb,amsfonts}%
\usepackage{amsthm}%
\usepackage{mathrsfs}%
\usepackage[title]{appendix}%
\usepackage{xcolor}%
\usepackage{textcomp}%
\usepackage{manyfoot}%
\usepackage{booktabs}%
\usepackage{algorithm}%
\usepackage{algorithmicx}%
\usepackage{algpseudocode}%
\usepackage{listings}%
\usepackage{lineno}

\begin{document}


\title{Interplay of Traditional Methods and Machine Learning Algorithms for Tagging Boosted Objects}

\author[1]{\fnm{Camellia} \sur{Bose}}
 
\author*[2]{\fnm{Amit} \sur{Chakraborty}}\email{amit.c@srmap.edu.in}

\author[3]{\fnm{Shreecheta} \sur{Chowdhury}}

\author[4]{\fnm{Saunak} \sur{Dutta}}

\affil[1]{\orgdiv{Center for High Energy Physics}, \orgname{Indian Institute of Science}, \orgaddress{\city{Bangalore}, \postcode{560012}, \country{India}}}

\affil*[2,3]{\orgdiv{Department of Physics}, \orgname{SRM University AP}, \orgaddress{\city{Amaravati}, \postcode{522240}, \country{India}}}

\affil[4]{\orgdiv{School of Science \& Technology}, \orgname{Vijaybhoomi University}, \orgaddress{\city{Greater Mumbai}, \postcode{410201}, \country{India}}}
%


\abstract
{
Interest in deep learning in collider physics has been growing in recent years, specifically in applying these methods in jet classification, anomaly detection, particle identification etc. Among those, jet classification using neural networks is one of the well-established areas. In this review, we discuss different tagging frameworks available to tag boosted objects, especially boosted Higgs boson and top quark, at the Large Hadron Collider (LHC). Our aim is to study the interplay of traditional jet substructure based methods with the state-of-the-art machine learning ones. In this methodology, we would gain some interpretability of those machine learning methods, and which in turn helps to propose hybrid taggers relevant for tagging of those boosted objects belonging to both Standard Model (SM) and physics beyond the SM. 
}


\maketitle
\section{Introduction}   \label{intro}

Despite the commendable triumph of the Standard Model (SM) in unifying the fundamental interactions \cite{Glashow:1961tr,Weinberg:1967tq, Salam:1968rm, Weinberg:2004kv} leading to the discovery of the Higgs boson at the Large Hadron Collider (LHC) experiment \cite{CMS:2013btf,ATLAS:2012yve}, several observational discrepancies cumulated from different experiments and theoretical inconsistencies encountered within the SM framework indicate itself as a low-energy effective theory of a formalism with higher symmetries \cite{Weinberg:1967tq}. With many theories Beyond the Standard Model (BSM) proposed, and many more to be formulated, it is a challenging task to figure out which of these theories explains the nature. Observations made at particle colliders which recreate the early universe with high energy density are presently the only trails to trace back the framework, most favored by the nature.

The BSM probes are, fundamentally, the classification problems: identification of the events that are governed by a BSM framework, among the ones accommodated in SM. Traditionally, the cut-based techniques have been the principal tool for the BSM probes. It identifies regions in the phase space with abundant (model dependent) BSM signatures over the SM backgrounds, and redirects the BSM search in those regions. With robust influx of collision data, constraining the phase space and pushing New Physics to higher energy frontier, implementation of advanced search strategies are inevitable. 

The focal point of attention in particle physics is presently the LHC \cite{Evans:2008zzb}. It offers a distinctive chance to investigate the dynamics of the SM at the TeV scale and to explore potential new physics signatures. A collision event at the LHC encompassing various objects, namely hadrons, leptons (electrons/muons), photons, neutrinos, and also other stable exotic particles associated to BSM physics if any. After reconstructing the leptons and photons using tracker and calorimeter information, the hadrons are clustered into jets which are the collimated sprays of particles resulting from the hadronization of quarks and gluons, along with their radiative effects. These are abundant in high-energy collisions, especially at the hadron colliders like LHC. Boosted objects, on the other hand, which are particles with high momentum relative to their mass, become more prevalent at higher energies. 
Therefore, analysis of jets and boosted objects at hadronic colliders provide great insights into the laws of nature. Over the recent past, an enormous amount of studies have been performed using the jet substructure method focusing on analyzing the physics with Higgs bosons and Top quarks at the highly boosted regime. Several ground-breaking advancements have been achieved in probing and tagging these objects, we refer \cite{Larkoski:2017jix} for a comprehensive review on this. Furthermore, proper identification and analysis of these boosted jets lead to an indirect probe of the properties of hypothetical new particles and potentially confirm or constrain the predictions of BSM theories. In other words, detecting such resonances in collider experiments can provide direct evidence for new physics beyond the SM. In view of anomaly detection, precise measurements of jet properties, such as their energy, angular distributions, and substructure, enables the comparison between experimental data with theoretical predictions based on the SM. Any deviations from these predictions could indicate the presence of new physics.

In parallel with technological advancement, the advent of Artificial Intelligence (AI) and Machine Learning (ML) algorithms have catered for the effective search strategies. ML algorithms are efficacious in carving out features pertinent for discerning BSM signatures, as deviations over the SM expectations. Contemporary machine learning techniques, encompassing deep learning, are finding application, adaptation, and advancement in the field of high-energy physics. The rapid progress in the field of Machine/Deep Learning and potential areas of application in High Energy Physics, especially in the context of tagging boosted objects using the complex and rich structure inside jets, has resulted in a renaissance. 

In this review, we aim to provide an overview of the Machine Learning tools and their implementations in BSM searches focusing on the tagging of Higgs bosons and Top quarks at the LHC. We will also provide an outline of the challenges ahead, and some of the issues/strategies need to be addresses in the near future. As a wide range of studies have been already performed, we will be somewhat selective in our presentation, and will choose a few representative studies. However, for interested readers we refer the following articles \cite{
Guest:2018yhq,Albertsson:2018maf,Radovic:2018dip,Bourilkov:2019yoi,Araz:2023mda} for modern dedicated reviews on this area, and also the Living Review of Machine Learning for Particle Physics \cite{Feickert:2021ajf} for a nearly comprehensive list of references which includes all the state-of-art techniques developed and applied in wide range of studies including the recent-most developments.

This work has been organized as follows: in Section \ref{sec:ANN} we provide a briefly outline of traditional machine learning algorithm, while relatively advanced variants of these techniques are discussed in Section \ref{sec:MLadv}. The main objective of this review is to discuss different studies on Jet tagging focussing on the Higgs boson and Top quark, which we discuss in Section \ref{sec:HiggsTop}. One of the major drawbacks of these machine learning models is the lack of interpretability. In Section \ref{sec:interpretable}, we highlight a few studies which focus on this domain and aim to capture relevant knowledge from a trained model in terms of the input features or high-level objects. In Section \ref{sec:summary}, we provide a summary along with the potential areas of improvement and directions to explore in future.


\section{Artificial Neural Network and Deep Learning: An Overview}   \label{sec:ANN}

Traditional Machine Learning tools provide a plethora of classification algorithms \cite{ClassRev1,ClassRev2} for efficient identification of BSM signatures from the background events. One such algorithm, the Decision Tree \cite{DT1}, functions in a way that resembles the traditional cut-based technique. This algorithm progresses by recursive imposition of cuts, on a given feature at a time, thereby splitting the phase space into subspaces based on the most significant attribute or feature \cite{DT2}, and finally providing a tree structure, constructed over the set of events, where each leaf node, attributed to a given subspace, represents a class label (signal or background). Decision trees are interpretable, easy to visualize and capture complex patterns leading to attribution of the class. Nevertheless they are prone to over-fitting \cite{DT3}. Ensemble learning techniques like Random Forest \cite{RF} and Gradient Boosting \cite{GrdBst1,GrdBst2} are implemented in improving the performance and robustness of decision trees. A Boosted Decision Tree (BDT) \cite{BDT1,BDT2,BDT3}, is a boosting technique that combines multiple decision tree stumps, to create a more robust and accurate predictive model. BDT trains the decision tree stumps sequentially over the set of events, such that the events misclassified in the preceding training are given more emphasis by the succeeding tree stumps. This process continues iteratively, with each tree stump focusing on the mistakes of the previous ones. The boosting technique is used to improve the performance of weak learners (the decision tree stumps) by combining their predictions.

In parallel, there has been advancement in development of ML algorithms by imitating the action of human brains, since 1940's \cite{ANN1,ANN2}, eventually leading to another paradigm of Machine Learning, the Neural Network (NN). A perceptron, that mimics the action of a neuron, the building block of nervous systems in intelligent species, was proposed in 1958 \cite{ANN3}, paving the way for the development of an architecture containing a single hidden layer of perceptrons \cite{ANN4} in 60's, termed Artificial Neural Network (ANN). It turns out that, a feed-forward artificial neural network \cite{FFNN1} with a single hidden layer containing a finite number of neurons can approximate any continuous function to any desired degree of accuracy, given a sufficiently large number of neurons \cite{UAT1,UAT2,UAT3}. In simpler terms, a neural network with just one hidden layer can approximate any function, as long as it has enough neurons in that hidden layer. This enables ANNs as universal function approximators, capable of representing a wide range of complex relations \cite{ANN5}. An ANN is trained upon fine-tuning the weights assigned between the neurons in adjacent layers, minimizing the error \cite{ANN6}, calculated during forward propagation, which is then used to adjust the weights iteratively \cite{ANN7} in the network, using the process termed back propagation \cite{BackProp1,BackProp2}. This involves computing the gradient of the error function with respect to the weights and updating the weights in a direction that minimizes the error \cite{GradDesc}.

The performance of an ANN architecture can be further enhanced by insertion of multiple hidden layers, leading to what is known as deep neural network (DNN) architectures \cite{DNN1,DNN2}. DNNs have accomplished even greater capacity for learning complex functions and patterns in the phase space that distinguish BSM signatures from SM backgrounds.

We will encounter the implication of these architectures and their variants in appropriate identification of simulated entities bearing BSM signatures in subsequent sections. The following section sketches an overview of advanced variants of DNN architectures.


\section{Advanced Variants of DNN: Autoencoders, Convolutional Nets, Transformers and Graph Network}  \label{sec:MLadv}

The BSM signatures manifest themselves as anomalies in the background of the SM events. A model independent search of New Physics is therefore, fundamentally focused on the effective identification of anomalies. Autoencoders are the simplest DNN architectures that can serve the purpose \cite{SciPstPhysCor,Cerri:2018anq}. An autoencoder possesses a bottleneck architecture \cite{pmlr-v27-baldi12a} of hidden layers of neurons, the bottleneck layer representing the compressed representation of the event features, while the preceding layers perform convolution and the following layers deconvolution of these features\cite{Hinton:AutoEnc}. It learns the efficient representations of events by training the network to encode input data into a lower-dimensional latent space and then decode it back to its original form. Thus, when trained with a large number of SM events, the autoencoder shall learn to reconstruct the background efficiently, by minimizing the reconstruction loss. BSM signatures, significantly different from SM counterparts, can be detected as events with high reconstruction loss by an autoencoder, trained over events governed by the principles of SM. Identified anomalies can be put for further validation and analysis using either traditional analysis methods or more sophisticated ML algorithms, to confirm whether they represent genuine BSM signatures or are experimental artifacts. Masked autoencoders, designed and adapted to efficiently learn from visual data, have been equally effective in jet-tagging from large-scale datasets of jet-images \cite{HeEtAl2021}.

Jet-tagging from jet images has been revolutionized with the implementation of Convolutional Neural Network \cite{O'Shea:2015}. This architecture learns hierarchical representations of features directly from the raw data, adapting to various jet signatures and complexities without the need for explicit feature engineering. With high scalability and inherent robustness to noises, CNNs exploit the property of spatial invariance for jets (\textit{i.e.}, the same convolutional filters are applicable across different regions of the jet image), to effectively capture local patterns and structures within jets, regardless of their position or orientation within the detector. The convolutional layer, the core building block of a CNN contains a set of learnable filters (termed kernels) that slide across the input image, computing the dot product between the filter and local regions of the input, leading to feature maps that highlight different aspects of the input image, such as edges, textures, or patterns \cite{CNN1}. We will explore the implementation of CNN in jet-tagging in the following sections.

The limitations of CNN in jet-image classification can be further overcome with the introduction of Transformers \cite{Vaswani:Transformer}. A transformer is a deep learning model adapted for processing of sequential data. Unlike CNNs, which typically rely on local receptive fields to capture spatial information, transformer models can capture long-range dependencies in the data more effectively, achieved through the self-attention mechanism, which allows the model to weigh the importance of different entries in a sequence during processing, thus allowing the model to capture long-range dependencies in the input sequence without being limited by the fixed-length context windows of CNNs. While CNNs rely on spatial convolutional operations to implicitly capture positional relationships, transformers explicitly model these relationships through the addition of positional encodings, which is important for tasks where the spatial arrangement of objects matters. On top, Transformers typically require fewer parameters compared to the CNN counterparts. This is because transformers process the input sequence as a whole, whereas CNNs apply convolutional operations locally, resulting in a larger number of parameters, especially in deeper architectures. As a result, transformers can achieve comparable or better performance with fewer parameters, leading to more efficient models. Transformer models also offer greater interpretability compared to CNN architectures, particularly in terms of understanding the relationships between different parts of the input sequence. The attention mechanism in transformers allows for visualization of attention weights, which indicate the importance of each input token or position for the final prediction, providing valuable insights into how the model makes decisions and which parts of the input contribute most to object identification.

We will conclude the section with an algorithm which has found extensive uses in event reconstructions \cite{
Thais:2022iok,Ehrke:2023cpn,Biscarat2021TowardsAR,Andrews:2018nwy,Andrews:2019faz} and particle identifications \cite{
ExaTrkX:2020nyf,Duarte:2020ngm,Qasim:2023pga}. The Graph Neural Network (GNN) is a variant of DNN, designed for analyzing graph-structured data. A Graph comprises of nodes (representing particles in a specific event) and edges (depicting relationships or connections between the nodes), with each node associated with a feature vector describing its properties (\textit{eg}. $p_T, ~\eta, ~\phi$) and its associated edges representing the respective parents and daughters. GNNs are used in reconstruction of the underlying objects (\textit{eg}. jets, leptons, photons) from the raw detector data obtained in colliders. Taking the detector outputs as input, GNNs cluster particles into jets \cite{Qu:2019gqs,Dreyer:2022yom}. GNNs identify individual particles with high efficiency and accuracy (\textit{eg}. electrons, muons, photons) based on their energy deposits and trajectories in the detector \cite{Komiske:2018cqr}. GNNs can learn to exploit subtle features and correlations in the detector data to improve the particle identification performance compared to traditional methods \cite{2024arXiv240102143L}. In addition, GNNs have been equally implemented in simulation of realistic collision events, essential for studying the performance of detectors, developing analysis techniques, and interpreting experimental results. GNN-based event generators can efficiently model the complex interactions of particles in the detector and provide samples that closely resemble real collision data.
A generalization over traditional graph structures, the hyper-graphs \cite{2020arXiv200205014O}, containing hyper-edges connecting more than two nodes provide additional flexibility of representing relationships among entities compared to traditional graphs. Hyper-graphs provide lucid representations of the complex interactions between particles in high energy collisions \cite{2024arXiv240103917C,2022arXiv220408770X}. Each node in the hyper-graph represents a particle, and hyper-edges represent interactions between multiple particles, enabling the analysis of particle decays, especially in situations where multiple particles are involved in the decay chain, along with the possible decay paths and their probabilities \cite{Shlomi:2020gdn}. From the robust data generated from the particle collider, hyper-graphs are used to analyze the topology of this data, identifying patterns and structures that may correspond to specific particle interactions or decay processes \cite{DiBello:2022iwf}.


\section{Identification of the Boosted Objects:}

\begin{figure}[!htb]
  \begin{center}
    \includegraphics[width=0.75\textwidth]{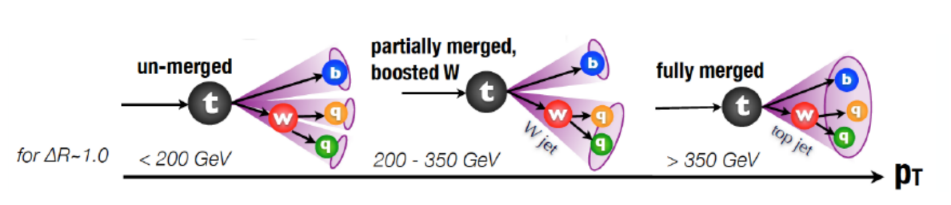}
  \end{center}
  \caption{A schematic diagram of the merger of top decay products with the increase of top quark $p_T$. The angular separation $\Delta R \sim 1$ has been used here. The figure has been adopted from \cite{elisethesis}. }  
  \label{fig:boostedjet}
\end{figure}

With technological advancement, as the collision energy progressively approaches the higher frontier, production of boosted objects \cite{Haller:2018mfm} becomes more abundant. A boosted object refers to a particle or a group of particles that carry a significant amount of momentum in a particular direction due to the high-energy collisions involved. Boosted objects typically arise in scenarios where the decay products of heavy particles, \textit{viz.}, top quarks, W or Z bosons, or Higgs bosons, are emitted with large momenta. Since these particles are produced with high momentum, their decay products tend to be collimated, forming a single, highly energetic jet of particles, encapsulating particles originating from the decay of the primary particle, with these decay products lying spatially close to each other within the detector. In Figure \ref{fig:boostedjet}, we provide a schematic diagram showing the way in which the decay products of a top quark merge with increasing $p_T$ of the mother particle.

The analysis of boosted objects has an advantage: the depletion of QCD Background at a high energy boosted regime, that enables enhanced signal density over the background. In parallel, several BSM theories lead to propositions of particles with higher resonances, eventually decaying to different stable states. Retracing such excited states requires effective reconstruction of these heavy resonances, necessitating the inclusion of their respective stable decay products in a single cluster. In a boosted regime, the stable states that originated from the decay of such a heavy particle can be enveloped in a single jet of large radius, known as large-R jets, enabling an opportunity to retrace the property of the intermediate resonance state.

There are several techniques to analyse boosted objects: traditionally, jet substructure algorithms are used to identify these patterns within a single large-radius jet. Substructure observables, namely, jet mass, N-subjettiness, energy correlation functions etc., are employed to distinguish between boosted objects from ordinary QCD background \cite{Larkoski:2017jix}. Boosted bosons like, $W, ~Z, ~H$, are identified from clustering their respective decay products and calculating their combined invariant masses after applying certain grooming techniques. Specific jet clustering algorithms, for example, Cambridge/Aachen or anti-kT clustering \cite{Salam:2010nqg} with a large radius parameter are often used to capture the entirety of the boosted object within a single jet and identify it. We refer Ref.\cite{Larkoski:2017jix} for a detailed discussion covering the basic principles underlying the calculation of jet substructure observables, the development of new observables leading to state-of-the-art machine learning techniques for jet substructure. 

The accuracy of boosted object identification has been enhanced with the implementation of ML techniques, stand-alone, or in association with the traditional approaches. Deep Learning models, trained on simulated data to recognize the characteristic features of boosted objects at colliders, are increasingly used in particle phenomenology for boosted object identification. Several tagging algorithms specifically designed to identify boosted objects, such as top quarks or Higgs bosons, have been formulated. These algorithms often exploit the unique properties of the boosted objects, such as their large transverse momentum or distinctive substructure. Particle flow algorithms enable effective reconstruction of individual particles from the collision debris, aiding in the identification of the boosted objects by accurately associating particles belonging to the same object. Overall event characteristics, such as missing transverse energy (MET), total transverse momentum, or other global event variables, can be analyzed to identify events containing boosted objects. 

These techniques are often used in combination to enhance the efficiency and reliability of boosted object identification in particle phenomenology experiments. Additionally, they are continuously evolving with advances in both experimental techniques and theoretical understanding. Below we discuss some of the recent-most developments for tagging the boosted Higgs boson and top quarks at the hadron collider environments such as the LHC.


\section{The Higgs and Top Tagging Algorithms}
\label{sec:HiggsTop}

\subsection{Higgs tagging }

Higgs boson, proposed by Brout, Englert and Higgs, is a neutral, spin-0 scalar accommodated in the SM to address the mass generation of the fermions and spin-1 gauge bosons after the breaking of electroweak symmetry. Probes carried out by LEP-2 collaboration in 2003 on Higgstrahlung ($e^{+} e^{-} \to HZ$) set the lower bound of the mass of Higgs boson to be, $M_H\geqslant$ 114.4 GeV \citep{LEPWorkingGroupforHiggsbosonsearches:2003ing}. However, the result was debatable for its assumption on high $HZZ$ coupling strength, and therefore could be nullified with BSM theories accommodating smaller $HZZ$ coupling strengths. Observation of a CP-even, spin-0, neutral scalar was finally made at LHC, separately by ATLAS and CMS collaborations in from the di-photon excess around the invariant mass of, $m_{\gamma\gamma} \sim 125.5$ GeV \cite{CMS:2013btf,ATLAS:2012yve}. The newly discovered particle was attributed to be the Higgs Boson, and today, Higgs has been observed through its (almost) all possible dominant decay modes. 

The pre-LHC colliders, \textit{viz.}, LEP, Tevatron, HERA provided the collision energies principally within the range between few 10's to 100's of GeV. The high energy collisions presently on the roll, at LHC, lead to abundant productions of heavy species (\textit{viz.}, $W$, $Z$, $H$ and $t$) with high transverse momentum. They are boosted, therefore their decay products are highly collimated along its axis of production and are observed in the calorimeters as a cluster of particles (abundantly, hadrons) within closer spatial vicinity between each other, as already discussed in previous section. After the discovery of the Higgs Boson at LHC in 2012 from its di-photon decay, it was difficult to probe the most abundant decay mode of Higgs to a pair of b quarks due to an insurmountable QCD background. Efforts were made to identify this decay mode, with initial advancements presented in the 2013 Moriond conference \citep{LHCHiggsconf:2013rie}, made over an integrated luminosity of 5 $fb^{-1}$ and 21 $fb^{-1}$ respectively for the collision energies, $\sqrt{s}$ = 7TeV and $\sqrt{s}$ = 8TeV. The measurement reached higher precision almost five years after the Higgs boson discovery, with the advancement of boosted b-tagging techniques \citep{ATL-PHYS-PUB-2015-035btag2,CMS:2017wtubtag3,Dhingra:2014nmabtag1}. Analysis of jet substructures for a boosted large-radius Higgs-jet turned out to be obvious in identifying the Higgs boson through collimated b-quark pairs. Below we start our discussion with the widely used BDRS Higgs tagger followed by recent-most developments employing Machine Learning methods to tag a boosted Higgs boson jet.

\subsubsection {The BDRS Tagger and its applications} 

The first successful attempt to probe a Higgs boson with mass around 120 GeV  at high transverse momenta employing state-of-the-art jet substructure technique was made at 2008 by Butterworth, Davison, Rubin and Salam \citep{Butterworth:2008iy_BDRS}\citep{Butterworth:2008iy_BDRS}. The approach proposed and followed in this work, popularly known as BDRS algorithm, termed after the authors, accomplished a remarkable improvement in Higgs boson tagging by accounting the jet substructures within a large-radius boosted Higgs-jet. The decay of the Higgs boson to a pair of b-quarks ($H \to b\bar{b}$) is considered, and the following steps are followed to identify a Higgs jet, 

\begin{enumerate}

\item The large-radius jet $j$ is dissected to obtain two subjets by reversing the final step of the clustering used to reconstruct the jet $j$. The two subjets, $j_1 $ and $j_2$ are labelled such that $m_{j_1} ~>~ m_{j_2}$.

\item In the context where a substantial mass drop is observed, \textit{i.e.}, $m_{j_1} < \mu m_j$ and the mass splitting between the two subjets is not large, (\textit{i.e} $\frac{min(p_{tj_1}^2, p_{tj_2}^2)}{m_j^2} \Delta R_{j_1,j_2} ^2 > y_{cut}$), assume $j$ to be the desired clustered object capturing the heavy particle and exit the loop. 

\item Else, redefine $j$ to be $j_1$ and restart the steps from 1.

\item Consider the final jet $j$ as the candidate Higgs boson only if both $j_1$ and $j_2$ are b-tagged.
\end{enumerate}

\begin{figure}[!ht]
  \begin{center}
    \includegraphics[width=0.75\textwidth]{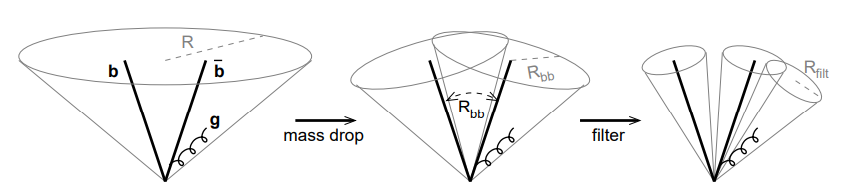}
  \end{center}
  \caption{Different stages of the BDRS Higgs tagging algorithm \cite{Butterworth:2008iy_BDRS}.}  
  \label{fig:BDRS}
\end{figure}

The two parameters $y_{cut}$ and $\mu$ are chosen such that the in addition to the dominant symmetric decay of Higgs boson to a pair of b-quarks, radiative decays of the final state quarks are also successfully captured within the large-R jet; the proposed analysis consider $\mu = 0.67$, and $y_{cut} = 0.09$. In reality, the above prescription is not yet optimal
for LHC, especially at the High $p_T$ regime. The primary reason for this is the significantly large contamination coming from underlying events (UE) due to the choice of a large jet radius. Therefore, a grooming technique, called Filtering, was proposed in addition to the above-mentioned procedure. 
The constituents of the large-R jet are re-clustered with a finer angular scale, $R_{\rm filt} < R_{b\bar{b}}$, and the three leading (in $p_T$) subjets are considered, among with the leading two subjets must be b-tagged. This last procedure helps in reducing the impact of UE events as well as improving the reconstruction efficiency. We summarize this procedure in Figure \ref{fig:BDRS}.  


The success of the BDRS algorithm in reconstructing  the Higgs boson jet (in fact the algorithm is very generic, it's applicable to any particle decaying to a pair of quarks) has found a plethora of applications in different studies involving SM as well as new physics particles. 

Even though jet substructure technique seems to be quite efficient in identifying and reconstructing the Higgs boson jets, the advent of machine learning algorithms has changed the whole landscape completely. The Higgs jet tagging methods driven by machine learning algorithms have achieved unprecedented success, showcasing promising performances, for example using Neural Network architecture in conjunction with N-Subjettiness variables \citep{Datta:2017lxt,Datta:2019ndh}, Convolutional Neural Networks and Graph Neural Networks \citep{Guo:2020vvt, Lin:2018cin, Li:2020grn, Alves:2019ppy,Choi:2023slq} and its variants, Interaction Network \citep{Moreno:2019neq}.

There has been progresses in other aspects of the study of Higgs boson using Machine Learning, which encompasses the effective identification of the production channel of boosted Higgs-boson from its substructure and event information \citep{Chung:2020ysf}, probing the magnitude and the phase of bottom Yukawa coupling ($y_b$) from the production of Higgs boson associated with $b \bar{b}$ \citep{Grojean:2020ech} with Decision Trees, probing BSM theories with extended Higgs sector through cascade Higgs decays to SM states using Long-Short Term Memory model \citep{Englert:2020ntw}, constraining the total decay width of Higgs boson from its inclusive production cross-section determined from modified Higgs jet reconstruction \citep{Harris:2019qwx}, and identifying the invisible decay modes of Higgs boson using DNN over low-level calorimeter observations \citep{Ngairangbam:2020ksz}.

Below we discuss a few studies on Higgs jet tagging using machine learning methods which found significant improvement over the existing Higgs tagging efficiencies. 

\subsubsection{Identification of Boosted Higgs using Two Stream CNN:}

An attempt to probe the Higgs bosons at the large $p_T$ regime through the dominant production and decay modes, namely production through the Gluon-fusion process and decay via $b\bar{b}$ final states, found to be very successful even though it receives a large contamination from the QCD backgrounds \citep{Lin:2018cin}. it's been achieved upon incorporating both the internal structure of the boosted Higgs jet and the global event features. The primary motivation to incorporate the global information comes from a couple of studies performed earlier which observed that there is additional information beyond traditional $n$-prong tagging methods for a boson which an image based analysis using CNN can capture \cite{deOliveira:2015} or can be estimated using a few simple observables \cite{Datta:2017lxt,Lim:2018toa}. In Ref.\cite{Lin:2018cin}, the authors therefore study the full potential of the Higgs tagging algorithm using all the available information in boosted Higgs events. 

\begin{figure}[!htb]
  \begin{center}
    \includegraphics[width=0.8\textwidth]{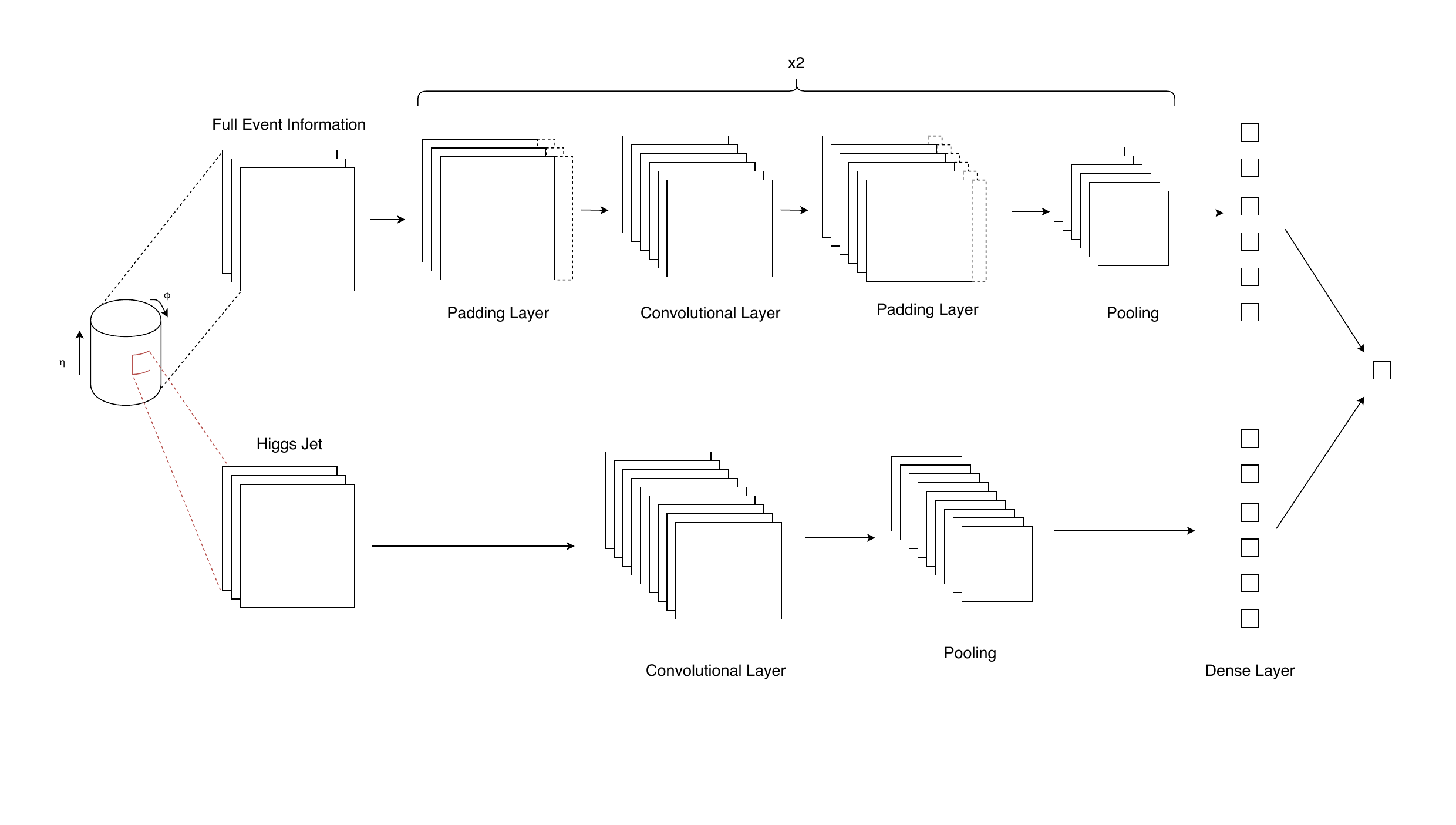}
  \end{center}
  \caption{A schematic diagram of the two-stream CNN used in \citep{Lin:2018cin}. The upper stream uses the full event information, while the bottom one relies on the jet substructure inputs.}  \label{fig:NN_architecture}
\end{figure}

The model deployed in \citep{Lin:2018cin} comprises of two CNN architectures functioning in parallel, as shown in Figure \ref{fig:NN_architecture}. While One layer analyses the information involving the double b-tagged substructures within the large-radius jet, the other layer processes the global event information. The network shown in Figure  \ref{fig:NN_architecture} contain the following details: size of each convolution filter is $5 \times 5$, the size of the pooling layers are $2 \times 2$, ReLu activation functions are used, along with a stride of 1. The first convolutional layer in each stream contains 32 filters, while the second convolutional layer possesses 64 of them. Each stream of the convolution networks flatten the respective output and passes the information through a dense layer of 300 neurons. Their respective outputs, carrying processed information on both the local and global traits, are finally transferred to the single neuron at the output layer of the combined two-stream architectures, equipped with sigmoid activation function, enabling the segregation of signal from the background.

\begin{figure}[!htb]
  \begin{center}
    \includegraphics[width=0.7\textwidth]{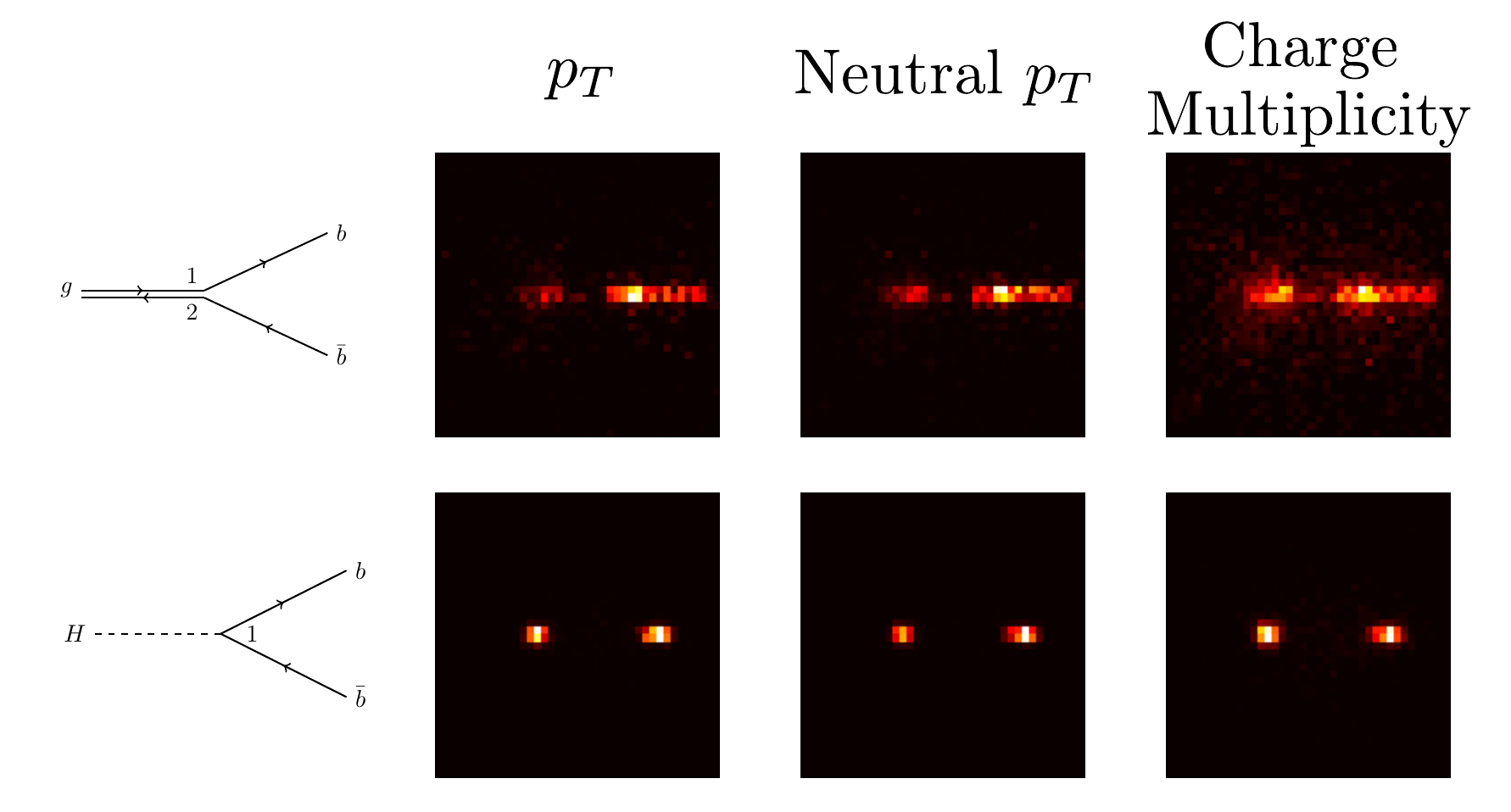}
  \caption{This figure, borrowed from \citep{Lin:2018cin}, represents effective differences in the three channels of a Higgs-jet from a QCD-jet. The large-radius jet of the color-neutral Higgs boson possesses a more contained color flow pattern than its QCD counterpart.}  \label{fig:jet-image-2cnn}
  \end{center}
  \end{figure}


Both the jet (local) and event (global) images include three types of information i.e., the pixel intensity of the images are defined in terms of the number of charged particles, $p_T$ of the charged particles, and $p_T$ of the neutral particles present in the jet/event. In Figure \ref{fig:jet-image-2cnn}, we show the average jet image for 100 background (top panel) and signal (bottom panel) events for the three above-mentioned channels. 
{From the images of the three channels, the two-prong structure of both signal and background jets is evident. The crucial difference lies in the radiation pattern and color connection between the two prongs constituted by the quarks and gluons. The Gluon being a colored object, the radiation is more spread out between the two prongs compared to the Higgs boson which is a color neutral object. The pixels between the two dominant prongs are therefore populated for the background jet unlike the signal jet. In order to encapsulate the color flow information, the authors of \cite{Lin:2018cin} used the variable $\beta_3$ calculated using the ratios of the N-subjettiness variable.} The novel method has achieved a maximum significance gain of ~2.2$\sigma$ against the signal efficiency of 25\%. Individually, the stream concerned with the local information on jet substructures contributes to a significance of ~2$\sigma$, and the one incorporating the global properties of the event touched a significance of ~1.4$\sigma$. As the numbers suggest, the local jet substructure based technique is indeed the most efficient approach, underscoring the notable contribution of global event information that can neither be overlooked.


\subsubsection{Machine Learning algorithms over Lund-Plane:}

Lund diagrams \cite{Andersson:1988gp} are theoretical representations of the phase space within the jets. They are constructed for individual jets from recursive declustering of the large-radius jet using Cambridge-Achen algorithm. The available phase-space, for a constituent at a given stage of declustering, in a Lund diagram, is mapped to a bi-dimensional logarithmic plane that portrays the transverse momentum and the angle of a given emission with respect to the emitter. Each subsequent emission leads to new phase space, represented as triangular leaf, for further emissions. They are particularly effective in analyzing parton showers and resummations and provide a powerful tool for visualization of the radiation taking within an individual jet \citep{Dreyer:2018nbflund}. This representation has found its robust usage in context of $W$-tagging \citep{Dreyer:2018nbflund}, later in top tagging \citep{Dreyer:2020brq}, and even for the identification of a charged species at LHC from embedding the tracker information on the phase space \citep{ATLAS:2020bbn}. Most importantly, Lund jet plane provides us the opportunity to gain insight on different underlying theoretical concepts which in turn improves classification performance. 

The success of Jet image technique in identifying and tagging boosted objects lead several studies improvising on the simple proposal, as already mentioned. One such attempt was made in \citep{lundplane}, where unprocessed jet images in (Primary) Lund plane were utilized to identify Higgs jets through a CNN architecture. Note that, Lund plane has the advantage in capturing the jet information in terms of the kinematic variables over the ordinary jet image method built using the calorimetric information in $\eta - \phi$ plane. The study analyzes the prospect of Higgs tagging through the $pp \to ZH$ process with Higgs decaying to a pair of b-quarks and Z boson decay to muons. The background jets are constructed using the same process except the Higgs decaying to a pair of gluons. The authors studied both moderate and high $p_T$ regimes of the Higgs boson.

\begin{minipage}{0.3\textwidth}
\includegraphics[width=0.5\textwidth]{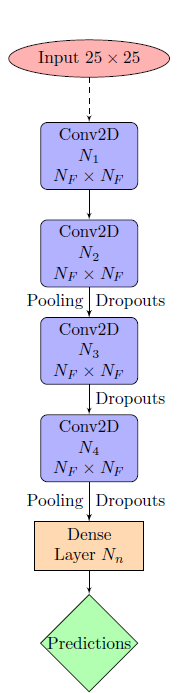}
\end{minipage}
\begin{minipage}{0.65\textwidth}
In Ref.\citep{lundplane}, the authors compare the performance of jet Lund Plane images with another well-established observable for tagging the color singlet particles, namely the jet color ring \cite{Buckley:2020kdpcolor_ring}. This observable is formulated through the comparison of signal and background matrix elements in the soft limit. Its construction ensures monotonicity concerning the likelihood ratio, establishing it as an optimal tagger under the provided approximations. Nevertheless, as indicated in \cite{Buckley:2020kdpcolor_ring}, the efficiency of the observable is found to be commendable for the $H \to b\bar{b}$, yet it proves insufficient when assessing the $H \to gg$ mode.
Using Lund jet images and color ring observable separately as input, a CNN architecture is designed, consisting of 4 convolutional layers followed by a dense layer of 800 nodes. The Max pooling layers after the second and fourth convolutional layers help in down-sampling the features. Additionally, to prevent overfitting of data, dropout of different strengths is applied after the second, third, and fourth convolutional layers. In each convolutional layer, a filter size of 3 is employed, but the number of filters differs for different datasets.  Throughout the network, the ReLU activation function is used for introducing non-linearity.
\end{minipage}

The study found that for $H \to b \bar{b}$ process with moderate boost, for 80\% signal efficiency, the background efficiency is around 20\%. Whereas at the high boosted regime, background efficiency is less than 10\% for the same signal efficiency. On the other hand, for the $H \to gg$ process with moderate boost, at 80\% signal efficiency, the background efficiency is around 40\%, which reduces to 15\% at the high boost regime. The ROC yields, as anticipated from \cite{Buckley:2020kdpcolor_ring}, the color ring variable has better efficiency in tagging the $H \to b\bar{b}$ mode than $H \to gg$. Comparison of the color ring observable to the Lund plane image concludes that it has a similar performance to the Lund plane image for moderate boost regime of $H \to b\bar{b}$ mode, while for the other three modes i.e., $H \to b\bar{b}$ in highly boosted regime and $H \to gg$ in both moderate and high boost regime, lund plane images outperforms color ring variables in the larger margin. Similar  analyses combining the high-level observables with the Lund jet plane were performed in Ref.\cite{Cavallini:2021vot}; the observations are similar to the one discussed above.  


\subsubsection{Graph Neural Network driven Higgs Tagging}

Graph Neural Networks (GNNs) constitute a class of deep learning techniques designed for obtaining inference on data characterized by graphs. These networks offer a simple and straightforward way to perform tasks involving node-level, edge-level, and graph-level predictions. They have found applications in mitigating pileup effects at hadron colliders, reconstructing trackers, identifying jets, classifying events, and study the interplay with traditional jet substructure methods; see for example \cite{ArjonaMartinez:2018eah,ExaTrkX:2020nyf,Qu:2019gqs,Mikuni:2020wpr,Ju:2020tbo,DeZoort:2023vrm, Mokhtar:2022pwm,Apresyan:2022tqw, Hammad:2023sbd}. Reconstruction of a large-radius Higgs-jet in boosted regime using a dynamic Graph Convolutional Network (GCN) architecture \cite{Wang:2018nkf} has been studied in \citep{Guo:2020vvt}. Each collision event, which is an agglomeration of detected particles characterized individually by $\eta,~\phi,~p_T,~E$ and a particle identity, is represented by a point cloud. The point cloud representation provides additional flexibility for the accommodation of all the particles, from a given event, in higher dimension feature space. 
Treating the point cloud as a graph, the network analyzes each event using K-Nearest-Neighbour algorithm to cluster individual node (representing a detected particle) with another node recursively, using the Euclidean metric between $\eta$ and $\phi$ of the pair of nodes under consideration. The architecture proposed in \citep{Guo:2020vvt} contains three EdgeConv blocks, with each block consisting of three multilayer perceptrons with 32, 64, 32 filters respectively. This is followed by two convolutional layers containing 32 and 2 kernels respectively, with a kernel dimension of $1 \times 1$. Over-fitting of the model is restricted by the introduction of a dropout layer with 25\% dropping rate. The final classification layer consists of neurons dotted with softmax activation function. The input feature consumed by the model is just the pseudo-rapidity and the azimuthal angle of the stable particles.
The model is trained for a sample containing Higgs boson with $p_T ~>~ 200$ GeV produced in association with jets, considering the Higgs decay to a pair of b-quarks. The training samples were overlaid with $\langle\mu\rangle = 50$ pileup events to account for the presence of multiple jets in each event. Special measures were taken to regularize the spatial coordinates of the boosted Higgs boson (in the particle-level) and its stable constituents. Additionally, the training sample was also purified from spurious backgrounds, by removing events with large-radius jet mass below 115 GeV. 

One of the key observations of this work is that the GCN driven analysis outperforms (by a factor of $\sim$1.5) the traditional jet substructure based analysis in the context of Higgs tagging efficiency as well as precision in momentum reconstruction. In fact, the sensitivity of pile up events to the overall performance of GCN is found to be negligible. It is noteworthy that the trained model is also capable to identify and tag the boosted Higgs jet in events of other production processes, however the performance of the tagger is degraded if additional boosted objects are present in the collision event in addition to the Higgs boson.  

For interested reader, we refer some of the recent-most studies where several state-of-the-art architectures based on GNN algorithms are proposed, for example \cite{Semlani:2023kzf,Aguilar-Saavedra:2023pde,Ehrke:2023cpn}. The tagging efficiencies have improved significantly in some of these cases.  

{Before we conclude this section, we would like to compare the performance of different Higgs taggers. In Table \ref{tab:Htagger}, we show the comparison of different architectures used for the Higgs tagging (first column), area under the curve (AUC) which is a measure of the ability of the classifier in distinguishing the signal over the backgrounds (second column), and the background rejection rate at 30\% signal efficiency (${\rm Rej}_{30\%}$). In the table, along with the previously discussed Higgs taggers, we also display the performance of the Double-b tagger widely used by the CMS collaboration at the LHC \cite{CMS:2020poo}. The performance of the Double-b tagger is comparable to some of the ML-based Higgs taggers discussed above. Note, we have estimated the background rejection rates from the available ROC curves. 
}
\begin{table}[!tbh]
    \centering
    \begin{tabular}{c|c | c}
    \hline
     & & \\ 
      Architecture   & AUC   & ${\rm Rej}_{30\%}$ \\ [2mm]
      \hline 
      BDT\cite{Alves:2019ppy} & 0.73 & 1.05 \\
      Resnet-50\cite{Alves:2019ppy} & 0.83 & 1.11 \\
      Two Stream CNN (ggF)\cite{Chung:2020ysf}  &  0.89 & 80 \\
      Two Stream CNN (VBF)\cite{Chung:2020ysf}  &  0.96 & 900 \\
      Graph Convolutional Network (GCN)\cite{Guo:2020vvt} & -  & 800 \\ 
       CS + LPCNN\cite{Cavallini:2021vot} & 0.89 & 25 \\ 
       CMS Double-b tagger \cite{CMS:2020poo} & - & 500 \\[1mm]
    \hline
    \end{tabular}
   
    \caption{Performance comparison between different Higgs tagging algorithms.}
    \label{tab:Htagger}
\end{table}

\subsection{Top Jet tagging }

Top quark is the heaviest candidate in the SM with a mass of ~172.2 GeV \cite{article:tawj}. It has a very short lifetime (~$10^{-25}$ s) \cite{Quadt2007TopQP}, much shorter than the time required for hadronisation \cite{Webber:1999ui,RevModPhys.82.2489,Grossman:2008qh,Humanic:2013xga}, and therefore decays to $b W$ with ~100\% branching ratio \cite{Deliot:2020ucq}, instead of undergoing hadonisation like other quarks. The $W$-boson thus produced may either undergo leptonic decay($W \to \ell \nu_\ell$), characterised by the registration of a charged lepton in the calorimeter and missing transverse energy, or hadronic decay ($W \to q \bar{q}$), manifested as localised deposition of, predominantly, hadronic energies at the calorimeters. The hadronic decay of $W$ is more abundant ($\mathcal{BR} \left(W \to q \bar{q} \right) ~67\%$) \cite{CMS:2022mhs}. When the top quark is produced with sufficient boost, the hadronic decay of $W$ leads to a large-radius top jet, comprising ideally of a 3-prong subjet-structures with an invariant mass close to the top-mass, and the invariant mass of two subjets, out of the three, around the $W$-mass. Such boosted large-R jets are crucial in different tests within the SM framework \cite{ALTARELLI:2005zv} and beyond \cite{Abdesselam:2010pt}. This section provides an overview of different techniques proposed for effective tagging of such boosted large-R top-jets.
The first top-tagger was proposed in 1994 \cite{Seymour:1993mx}, way before the discovery of top at Tevatron in 1995 \citep{topdiscov1,topdiscov2}. This top-tagger, based on a recombination algorithm, achieved better top-tagging efficiency over the pre-existing cone algorithms \cite{Andrieu:2005re,Seymour:2000yy}. %
The topjet was identified from the leptonic decay mode of $W$, assuming the top-mass $\lesssim$ 200 GeV. Events with top-pair production from $p \bar{p}$-collision ($p \bar{p} \to t \bar{t}$) were considered, followed by semi-leptonic decay of the top ($t \to b W, ~ W \to \ell \nu_\ell$). This resulted in the observation of events with four jets, two originated from the b-quarks and two from the charged leptons, and missing transverse energy, along with few possible QCD jets due to radiative effects of b-hadrons. This tagger did not employ any b-tagging algorithm to identify the b-jets originated from the top-decay. There have been several improvisations, \citep{Ellis:2009me_tagger, Butterworth:2002tt_tagger, Ellis:2009su_tagger, Plehn:2011sj_tagger} proposed over the first top tagging algorithm, addressing both the hadronic and semi-leptonic\citep{Chakraborty:2023dhw, Chatterjee:2019hyk} decay channels of top.
 
\subsubsection{Traditional Top taggers}

The first public top tagger with BDRS setup was the Johns Hopkins top tagger\citep{JHTop_tagger}, which considers the top decaying hadronically in a highly boosted regime, with a minimum transverse energy ($E_T$) of 1 TeV and the $p_T$ accounting for 35\% of its $E_T$, constructed with Cambridge-Achen jet clustering algorithm. The radius of the large-R top-jet for this tagger is set to be 0.8, and jet substructures carrying at least 10\% of the large-R jet $p_T$ inside the large-R jet with a minimum angular separation of 0.19 are determined following sequential de-clustering of the parent jet. The process terminates once the clusters corresponding to the $W$ and the top from two successive clustering are identified, imposing the following criteria: \\

  \begin{enumerate}
  \item The invariant mass of the three subjets must lie within a window of 30 GeV of the top-mass, $$ m_{jjj} ~=~ m_{t} ~\pm 30\, \mathrm{GeV.} $$
  \item The invariant mass of one of the pairwise combinations of the subjets must lie within a window of 15 GeV of the $W$-mass, $$  m_{jj} ~=~ m_{W} ~\pm 15\, \mathrm{GeV.} $$
  \item The helicity angle of the top, measured in the rest frame of the reconstructed $W$-boson is small, $$ cos\theta_h ~<~ 0.7. $$
  \end{enumerate}
The reconstruction of the $W$-boson from its hadronic decay, in general, is overshadowed by the predominant existence of the QCD background, which peaks around 50 GeV and depletes slowly with a flatter tail. This tagger, therefore, in order to minimise the QCD background from the $W$-resonance window, focuses on high energy boosted regime of the top-production, thereby further narrowing the available phase space and consequently the cross-section for the top production for the energy reach of the contemporary colliders. Hence there were propositions with certain modifications that could relax the constricted phase space the JHTopTagger demands with minimal compromise over the QCD background. 

The CMS toptagger \citep{CMS:2009lxa_tagger}, which focuses on large-R jets with identical jet radius but with minimum transverse momentum, $p_T^{min} ~=~ 250\text{ GeV}$, follows similar declustering process and imposes:

\begin{itemize}
\item Pair of jet substructures emerged within the large-R jet, after reversing the final clustering step, must carry at least 5\% of the large-R jet $p_T$, $p_T^{\rm sub-cluster} > 0.05 \times p_T^{\rm fatjet}$ individually. The pair is retained and the rest of the cluster is removed, to eliminate the low-energy subclusters, which originated mainly from the radiative effects within the large-R jet.
\item If only one such substructure is obtained satisfying the above condition, then the above approach is repeated on the said sub-structure until one obtains a pair of them, satisfying the lower bound on the transverse momentum mentioned above. Else the large-R jet is no longer taken into consideration.
\item Once the substructure pair is determined, an identical approach is made to de-cluster them individually, satisfying the same $p_T$ constraint. After the completion of this step one either obtains four jet sub-structures carrying at least 5\% of the large-R jet $p_T$, or three subjet structures followed from absence of a hard cluster within one of the two parent sub-structures.
\item The invariant mass of these maximum of four subjet structures must satisfy, $100\text{ Gev} \leq m_{\rm jet} \leq 250\text{ Gev}$ and the minimum of the pair-wise invariant mass of all possible combinations of these subjets must be greater than 50 GeV.
\end{itemize}

HEPTopTagger \citep{Plehn:HEPTOPTagger}, on the other hand, further relaxes the minimum $p_T$-cut over the boosted large-R jet. It considers a large-R jet with $p_T^{min} ~=~ 200\text{ GeV}$ and radius $R = 1.5$, optimised to incorporate the stable hadronised states formed from the three principal decay products of the top quark ($b,q,q'$), keeping an option to implement b-tagging to further reduce the QCD background. The tagging algorithm advances similar to the BDRS algorithm, in the following fashion:
\begin{itemize} 
\item The first de-clustering step leads to two subclusters, labelled $j_1$ and $j_2$, such that $m_{j_1} > m_{j_1}$. If $m_{j_1} < 0.8 m_j$ ($j$ refers to the jet), then both the substructures are retained, else $j_2$ is rejected from further consideration.
\item For each jet substructure $j_i$, the sequential de-clustering is iteratively implemented till further de-clustering leads to subclusters of mass $<~ 30$ GeV. This terminates the de-clustering process.
\item Once all the subclusters are obtained, jet filtering is performed to obtain subjets with size, $R = 0.3$, retaining at most five hardest subjets. Next, all possible combinations of three out of five subjets are considered and their invariant masses are determined. The combination with invariant mass closest to the top-mass is retained for further considerations.
\item Of the three subjets finally retained, labelled $j_1$, $j_2$ and $j_3$, pairwise invariant masses are calculated $m_{12}$, $m_{23}$ and $m_{31}$. The candidate large-R jet is top-tagged if one of the following conditions are satisfied:
\begin{itemize}
\item $ 0.2 < \tan^{-1} \left( \frac{m_{13}}{m_{12}} \right) < 1.3 \qquad \text{and} \quad R_{\min} < \frac{m_{23}}{m_{123}} < R_{\max} $, ($R_{\min} = 85\% \times \frac{m_W}{m_t}$, $R_{\max} = 115\% \times \frac{m_W}{m_t}$)
\item $ R_{\min}^2 \left( 1 + \left( \frac{m_{13}}{m_{12}} \right)^2 \right) < 1 - \left( \frac{m_{23}}{m_{123}} \right)^2 < R_{\max}^2 \left( 1 + \left( \frac{m_{13}}{m_{12}} \right)^2 \right)  	
\quad \text{and} \quad \frac{m_{23}}{m_{123}} > 0.35 $
\item $ R_{\min}^2 \left( 1 + \left( \frac{m_{12}}{m_{13}} \right)^2 \right) < 1 - \left( \frac{m_{23}}{m_{123}} \right)^2 < R_{\max}^2 \left( 1 + \left( \frac{m_{12}}{m_{13}} \right)^2 \right)  	
\quad \text{and} \quad \frac{m_{23}}{m_{123}} > 0.35 $
\end{itemize} 
\end{itemize}

\begin{figure}[!htb]
\centering   
\includegraphics[width=.3\textwidth]{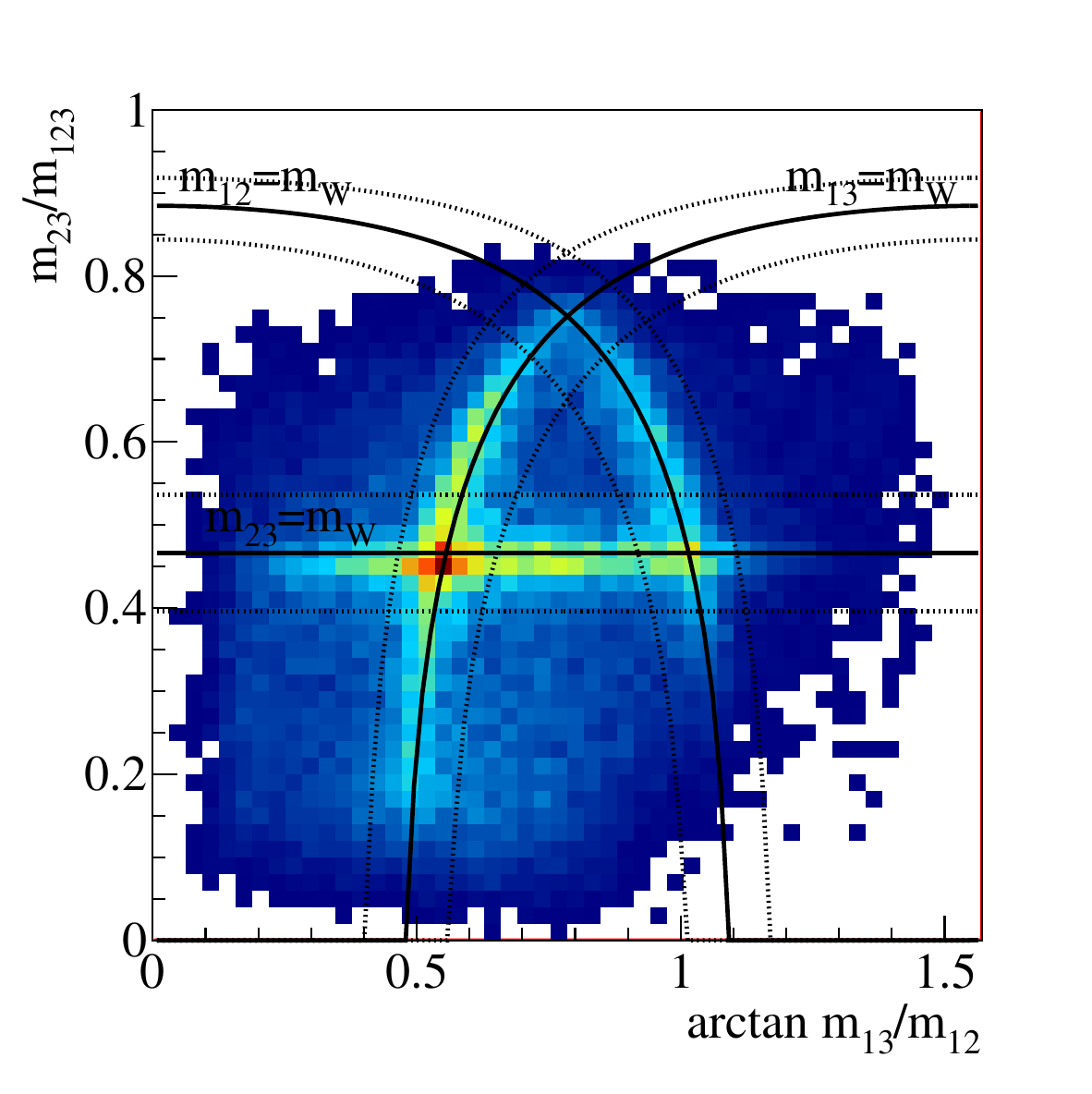} 
\includegraphics[width=.3\textwidth]{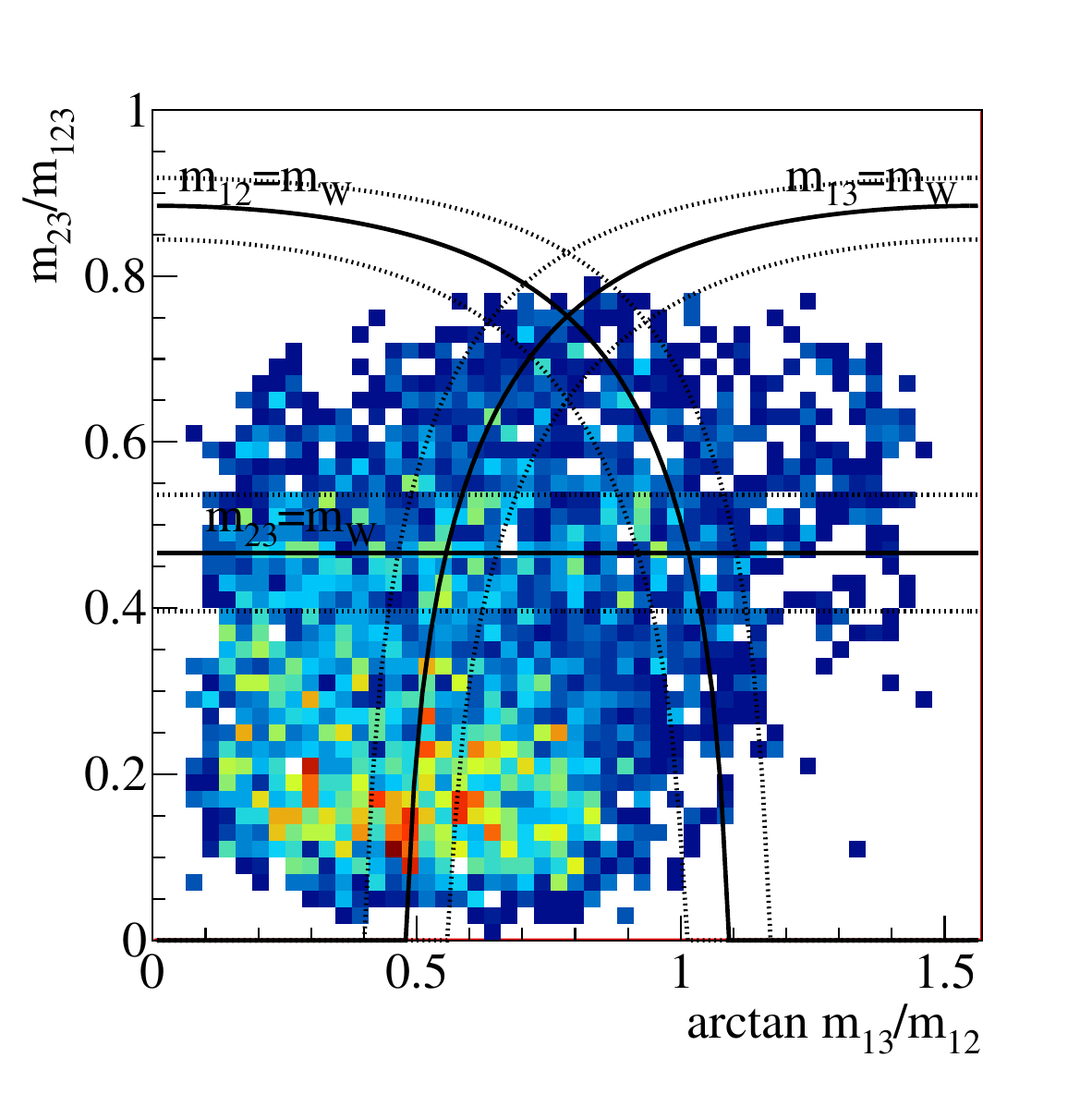}
\includegraphics[width=.3\textwidth]{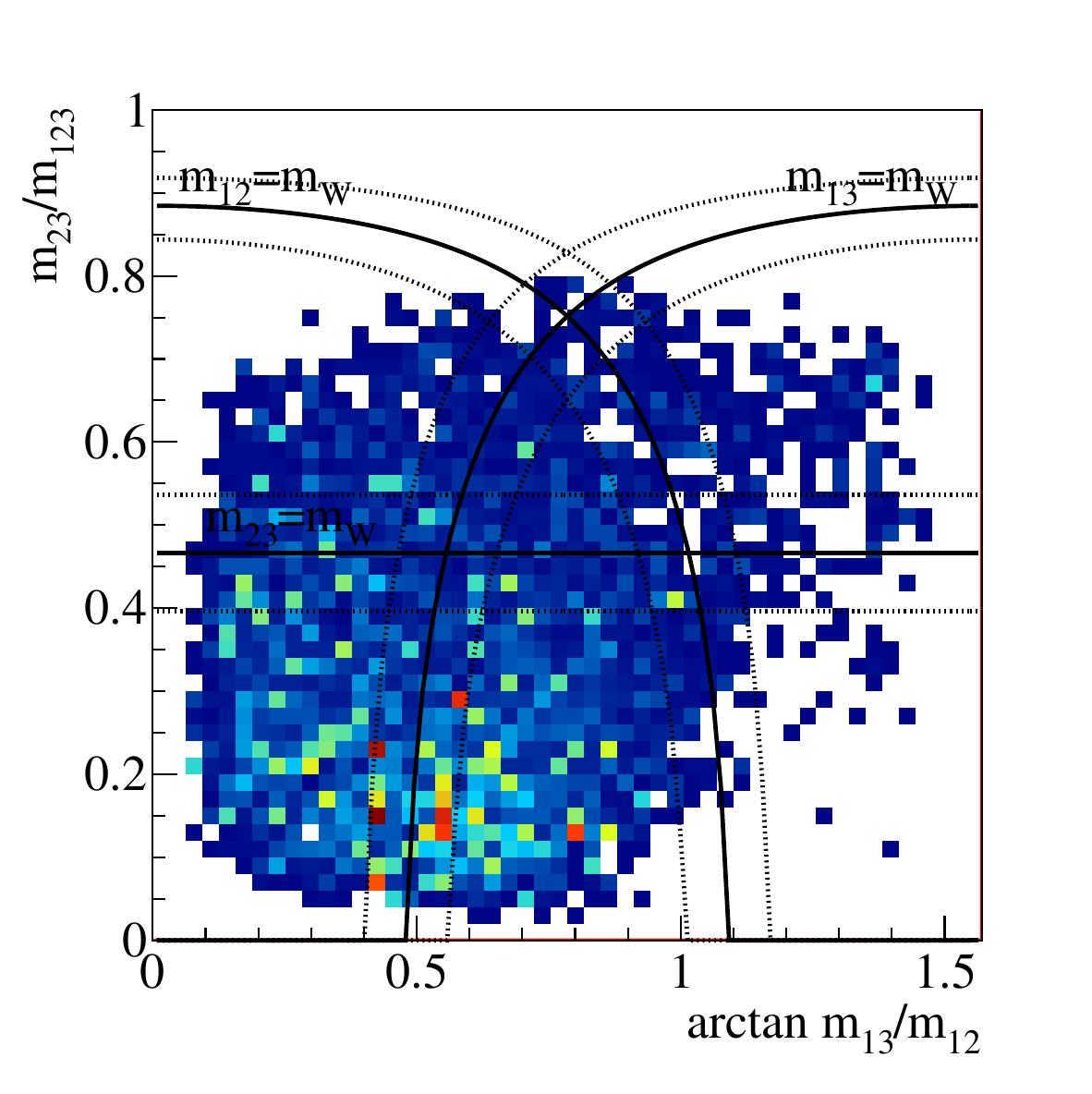}
\caption{The distribution of events in the $\arctan \left( \frac{m_{13}}{m_{12}} \right)$ vs $\frac{m_{23}}{m_{123}}$ plane for $t\bar{t}$, $W$-jet and QCD events \cite{Plehn:HEPTOPTagger}. The region in red shows comparatively higher density of events.} \label{Fig:massplane}
\end{figure}

Figure \ref{Fig:massplane} shows the effective discriminators, $\frac{m_{13}}{m_{12}}$ and $\frac{m_{23}}{m_{123}}$ which efficiently dissect portions on the phase space with abundant top-jet events from $W$ and QCD jets.

There have been several other traditional top-tagging algorithms, interested readers may look into \cite{Plehn:2011tg_tagger} for ready references.

Despite the commendable success these traditional approaches accomplish, advent and implementation of Machine Learning techniques have enhanced the top tagging efficiency remarkably. Propositions of effective tagging algorithms involving different variants of Neural Network architectures, namely, CNN \citep{deOliveira:2015xxd,Chakraborty:2020yfc,Qu:2019gqs}, GNN \citep{Shlomi:2020gdn_GNN, Dreyer:2020brq}, Autoencoders \citep{Finke:2021sdf_Autoencoder} and Transformers \citep{Qu2022Part_trans} have revolutionized the top-tagging strategies. Below we introduce some of these state-of-the-art taggers. 


\subsubsection{Multi Layered Perceptron and top-tagging} 
Multi Layered Perceptron (MLP) or Deep Neural Network composed of multiple hidden layers of perceptrons has been discussed in Section \ref{sec:ANN}. With densely connected architecture,they are effective in determining the complex relation between the features that tells apart a given jet flavour from the rest. DNN has been successfully implemented in \citep{Pearkes:2017hku} for the identification of a boosted top-jet. The architecture incorporates 4 densely connected hidden layers, containing respectively 200, 102, 12 and 6 perceptrons, with each layer using ReLU activation function to counter vanishing gradient problems during the model training. The output layer consists of a single perceptron with sigmoid function to distinguish top-jets from the rest. Inputs, which are the observables related to the candidate large-R jet, are given into the network as vectors for training and validation. The model achieved a signal efficiency of 50\% against a background rejection rate of 45\%, for the top samples with $p_T$ $\in$ [600, 2500] GeV. The performance of such architecture enhances two-fold compared to the cut-based taggers, with a DNN dotted with densely connected 5 hidden layers using ReLU activation function and containing 18, 16, 14, 10 and 5 perceptrons respectively \citep{ATLAS:2017jiz}, for the top-candidates with $p_T ~\in~ [400,~2000]$ GeV. We will conclude the discussion with DeepTopLoLa tagger \citep{Butter:2017cot}, which consumes the Lorentz vectors of jet constituents as inputs. This tagger comprises of a combination layer that emulates QCD-inspired jet recombination, a Lorentz layer translating the 4-vectors into relevant kinematic observables, and two fully connected hidden layers. Using the calorimeter and tracker information at full experimental resolution, the tagger turns out to be effective in the highly boosted regime, matching with the performance of image based CNN tagging algorithms. 


\subsubsection{RNN based top-tagger}
\begin{figure}[!htb]
\centering
\includegraphics[width=0.5\textwidth]{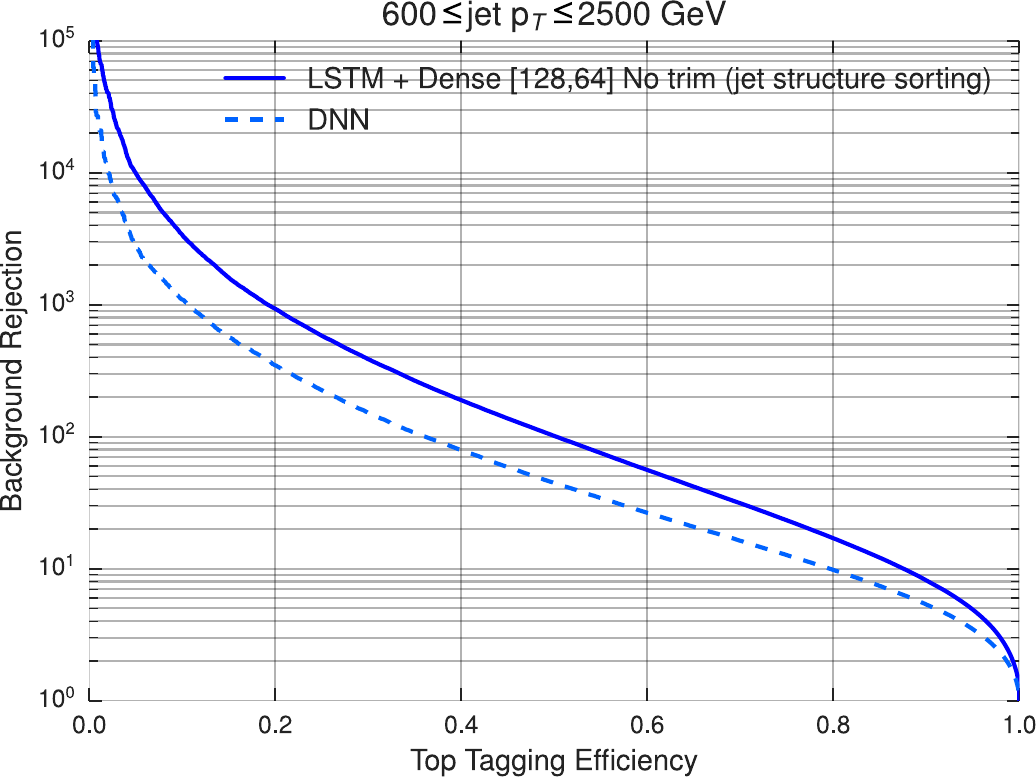}
\caption{The ROC curve depicting the top tagging efficiency for the RNN based top tagger used in \citep{Egan:2017ojy}.}  \label{fig:RNN}
\end{figure}

Recurrent Neural Network (RNN) is designed to process sequential data, with recurrence of the input sequence used by the successive layers of the architecture \citep{RNN1,RNN2,RNN3,RNN4}. This leads to recursive imposition of a single weight, fine tuned by the gradient descent, which in turn ends up at vanishing/exploding gradient problem \citep{2012arXiv1211.5063P}. Long Short-Term Memory(LSTM) models bypass this problem by truncating the gradient \citep{LSTM1}, and assign non-trivial weights at individual inputs in the sequence. This algorithm has been implemented in tagging boosted top-jets from their respective constituents \citep{Egan:2017ojy}. The architecture comprises of an LSTM layer with a state width of 128, followed by a fully connected dense layer containing 64 perceptrons. Adam optimizer, used in the model provided the most stable training with best performance over the unseen dataset. Large-R jets with $p_T$ $\in$ [600, 2500] GeV and $\lvert \eta \rvert ~\leq~$2.0 were considered. The jet substructure information, \textit{i.e.}, $p_T$, $\eta$, $\phi$, along with the subjet constituents were taken as model input. The tagger achieved better performance than a DNN-based tagger \citep{Pearkes:2017hku} across all signal efficiencies, accomplishing 100\% background rejection rate for the signal efficiency of 50\%, as presented in Figure \ref{fig:RNN}.

\subsubsection{CNN based top-tagger:}
Convolutional neural network(CNN) is a deep learning architecture, predominantly used for object identification from respective images. CNN has a noteworthy contribution in particle physics phenomenology, particularly in recognition of particle tracks \cite{Komiske:2018cqr}, characterization of rare decay events \cite{Baldi:2014kfa_cnn_rare}, identification of of jet flavours through jet images \citep{CMS:2020poo}. 
CNN has been a rapidly developing paradigm of Deep Learning and has witnessed remarkable breakthroughs in layer design, regularizations, optimizations and swift computations \citep{2015arXiv151207108G}. It has proved to be very effective for tagging top jets in both the high and low $p_T$ regimes. The pioneering work on incorporation of Computer Vision techniques in jet images analysis was done in \citep{Cogan:2014oua}. Constructing the image of a given jet from the calorimeter tower information, Fisher discriminant analysis was implemented to obtain a discriminant that could tell apart jet images obtained from hadronic decay of boosted $W$-boson from the QCD background. The prescription followed in \citep{Cogan:2014oua} accomplished better discrimination power over the contemporary substructure approaches, additionally providing significant insight into the internal structure of jets. 

An extended version of this work has been carried forward in \citep{deOliveira:2015xxd}, identifying boosted $W$-jets with $p_T ~\in~ [250,~300]$ GeV, reconstructed using anti-$k_T$ algorithm from respective hadronic decays. The large-R $W$-jet was first trimmed by reclustering its constituents into $R$ = 0.3 with $k_T$ algorithm, and dropping those subjets with $p_T^{subjet} ~<~ 0.5 \times p_T^{jet}$. The jet image is constructed pixel-by-pixel from the energy deposition of its constituents on the $\left( \eta, \phi \right)$-plane, by matching the pixel-size with the granularity of the calorimeter tower, and the intensity of each pixel determined by, either of, the total energy deposition of all jet constituents inside the concerned $\left( \eta, \phi \right)$-pixel, or the transverse projection of the jet energy in the tower corresponding to the spatial coordinates, $\left( \eta, \phi \right)$. The pre-processing of the image thus obtained was performed in four steps: 
\begin{enumerate}
\item Translation of jet images such that the leading subjet lies at $\left( \eta = 0, \phi = 0 \right)$. In order to preserve the pixel intensity under translation, transverse momentum replaces the transverse energy for reconstituting jet images. 
\item Rotation of the jet image around its center, such that the second leading subjet, if exists, is at $-\pi / 2$. Else, the image is so rotated that the first principle component axis of the pixel intensity distribution is aligned vertically.
\item Re-pixelation of the rotated image, followed by the redistribution of the energy deposits in the rotated grid using a cubic spline interpolation, such that the rotated grid is aligned with the original grid.
\item Inversion of the jet image, resulting from a parity flip, such that the right side of the jet image possesses the highest sum pixel intensity.
\end{enumerate}
The CNN architecture consists of total 14 hidden layers: 3 sequential units, each consisting of a convolution layer, a Max-Pool layer and a DropOut layer, followed by a Local Response Normalization layer and two fully connected dense layers. The processed information finally reaches the output layer containing a perceptron with sigmoid activation to identify the $W$-jets from QCD counterparts, through another DropOut layer. All three convolution layers use 32 filters, with dimensions, $11 \times 11$, $3 \times 3$ and $3 \times 3$ respectively, regularised with the $\mathcal{L}^2$ norm. Three MaxPool layers perform down-sampling of (2, 2), (3, 3), and (3, 3) respectively and all three dropout layers encountered in the sequential units uses a dropout of 20\%, while the one preceding the output layer applies the same amounting to 10\%. The dense layers individually carry 64 perceptrons.
%

\begin{figure}[!htb]
\centering   
\includegraphics[width=.45\textwidth]{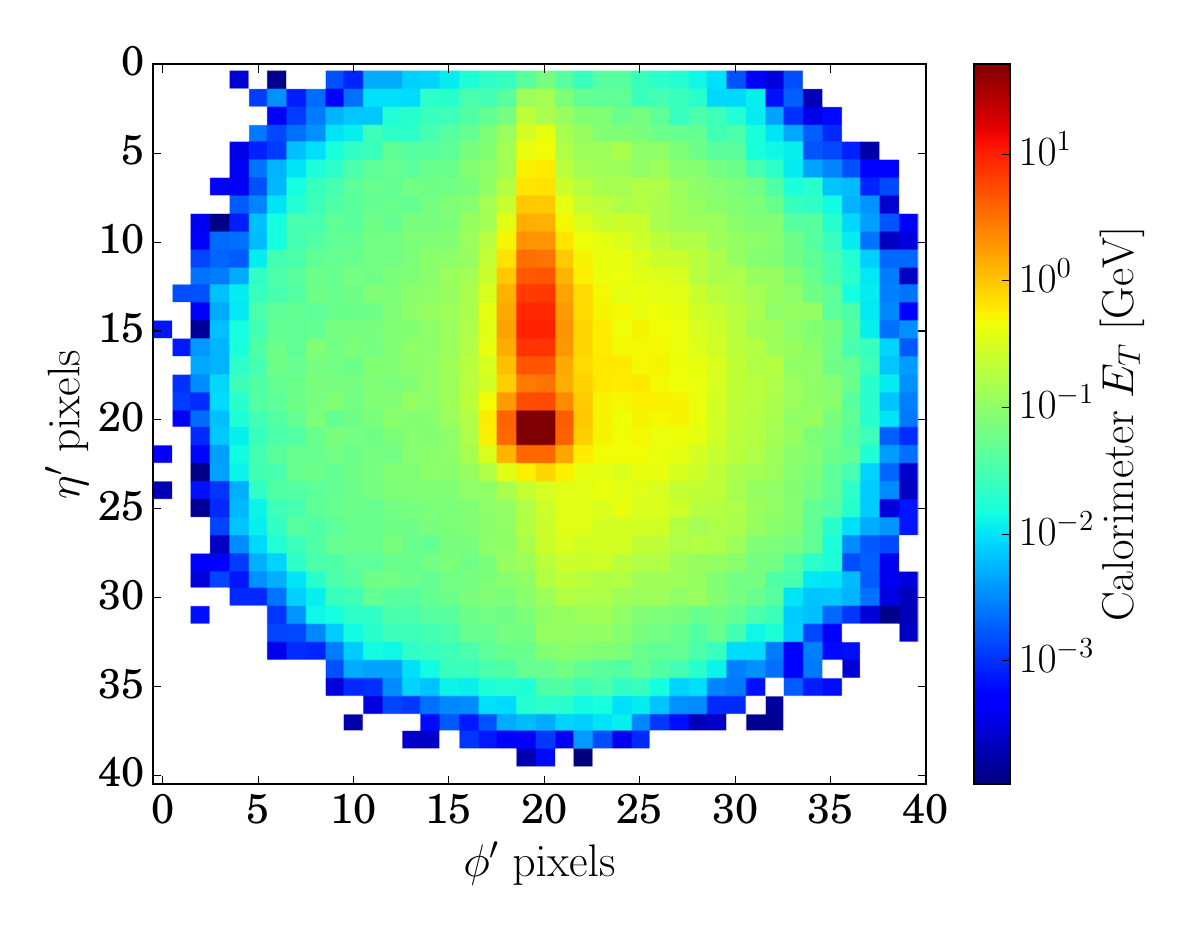}
\includegraphics[width=.45\textwidth]{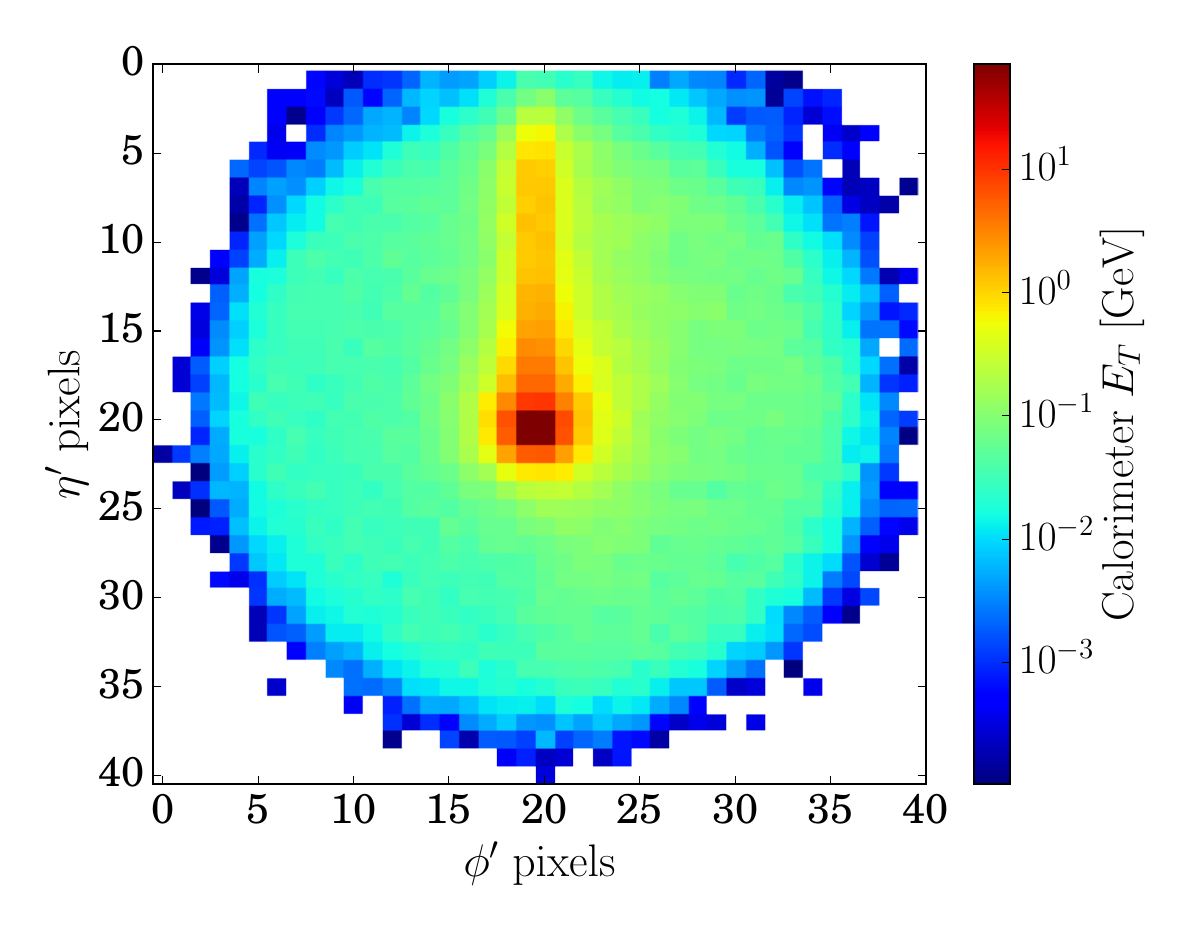} 
\caption{Figure borrowed from \citep{Kasieczka:2017nvn_deeptop} portraying the top and QCD jet-images used as input. Left panel: Image of the top candidate jet with three prong structure, with the leading $E_T$-ordered subjet lying at the center, the sub-leading one at 12 o'clock position, and the third one to the right of the leading subjet. Right panel: Image of the background QCD jet with the leading subjet at the center. } \label{Fig:jet-image}
\end{figure}

\begin{figure}[!htb]
\centering   
\includegraphics[width=.8\textwidth]{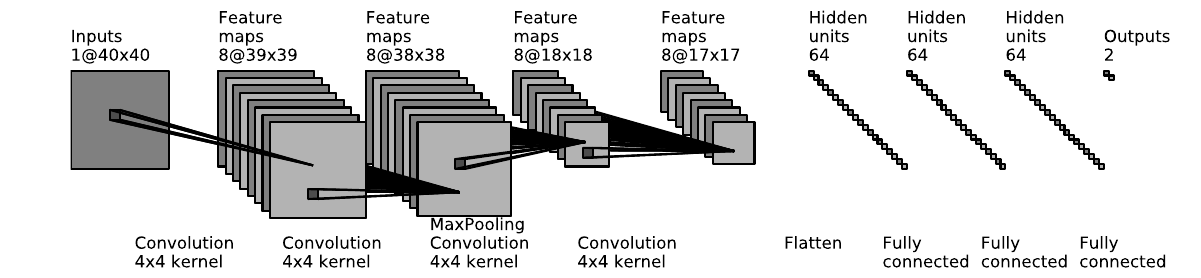}
\caption{Detailed outline of the CNN architecture of DeepTopTagger, proposed in \citep{Kasieczka:2017nvn_deeptop}.} \label{Fig:CNNarchDeepTop}
\end{figure}

%
The application of CNN for tagging top-jets was introduced in \citep{Kasieczka:2017nvn_deeptop}. Jet images were preprocessed following the steps prescribed in \citep{deOliveira:2015xxd} and were cropped into dimensions of $40 \times 40$. Figure \ref{Fig:jet-image} presents the jet images of a top-candidate and a QCD jet after the image preprocessing.
The model architecture proposed and trained in \citep{Kasieczka:2017nvn_deeptop} is shown in Figure \ref{Fig:CNNarchDeepTop}. Each $40 \times 40$ jet-images were passed through two successive CNN Blocks, each implementing filters with dimensions $4 \times 4$ and stride 1, to construct 8 Feature Maps, of dimensions shorter by 1 compared to that of the preceding input. The output of 8 Feature Maps of dimensions $38 \times 38$, thus obtained, were next passed through a Max-Pool layer followed by a CNN layer to create 8 Feature Maps of size $18 \times 18$, and through another CNN layer to reduce them to $17 \times 17$. The resultant output is then flattened, and propagated through three dense, fully connected hidden layers, each containing 64 perceptrons, and finally to the output layer containing a single perceptron dotted with sigmoid function to classify top-jets from the QCD background.
The performance of the model thus proposed, when compared to that of a conventional, multivariate QCD-based top-tagging using standard large-R jets, infers the convolutional networks a prospective novel approach for multivariate hypothesis-based top tagging. The work has been further extended in \citep{Macaluso:2018tck} where the same architectural set up, as in \citep{Kasieczka:2017nvn_deeptop}, was followed. Increasing the number of feature maps to 128, and perceptrons in the dense layers to 256, an overall improvement of a factor of $\sim$ 10 is achieved over the preceding model \citep{Kasieczka:2017nvn_deeptop}.

There have been further studies on CNN based tagging, on $30 \times 30$ pixelated images constructed from HCAL \citep{Choi:2018dag}, with each pixel representing energy deposits of the jet constituents. The architecture  incorporates a single convolution layer comprising of 30 filters followed by a dense layer leading to the output layer with a single perceptron dotted with sigmoid activation segregating the signal from the QCD background. The CNN defined observables for the tagger are IRC safe as long as the gluons are non-collinear with all the quarks inside the jet, along with the performance of this tagger being resilient to the kind of signal used, as established using two different signal samples, one as boosted top quark and the other with a gluon in the final state.

The tagger proposed in \citep{Bhattacharya:2021ghu} works for both hadronically and leptonically decaying tops in high $p_T$ regime. The architecture proposed in \citep{Bhattacharya:2021ghu} consists of 3 convolution layers with $10 \times 10$, $5 \times 5$ and $2 \times 2$ kernels respectively, with each convolution layer containing ReLU activation function followed by a MaxPool layer of stride $2 \times 2$ with $2 \times 2$ kernels, and finally flattening the output before propagating through 2 dense layers with respectively 90 and 50 perceptrons along with ReLU and Sigmoid activation functions. The tagger is advantageous in the sense that it does not require lepton identification to tag leptonically decaying top events. It turns out to perform slightly better with the leptonically decaying sample than the hadronic ones. 


\subsubsection{GNN based top-tagger:}
\begin{figure}[!htb]
\centering
\includegraphics[width=0.2\textwidth]{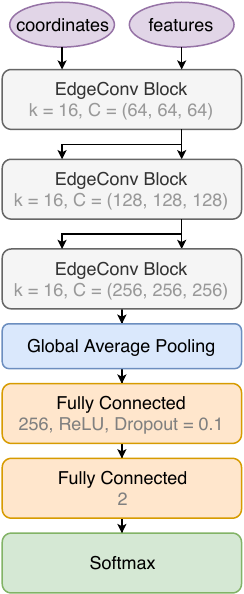}
\caption{Figure portraying the ParticleNet architecture employed in  \citep{Qu:2019gqs}.}  \label{fig:ParticleNet}
\end{figure}

The Graph Neural Network has been an emerging tool for tagging jet flavours. The ParticleNet algorithm, using GNN architecture, is a recent advancement in top tagging, with the jets treated as an unordered collection of particles, termed particle cloud \citep{Qu:2019gqs}. The architecture of ParticleNet algorithm is presented in Figure \ref{fig:ParticleNet}. 

The algorithm of ParticleNet rests on the notion of Dynamic Graph Convolutional Neural Network (DGCNN): 3 EdgeConv block with (64, 64, 64), (128, 128, 128), and (256, 256, 256) channels respectively is employed seeking 16 nearest neighbors, followed by a global average pooling layer aggregating all the learned features of the constituents of particle clouds. Thereafter 2 fully connected dense layers along with a single perceptron containing the output layer completes the architecture and generates binary classification output. 

\begin{figure}[!htb]
\centering
\includegraphics[width=1\linewidth]{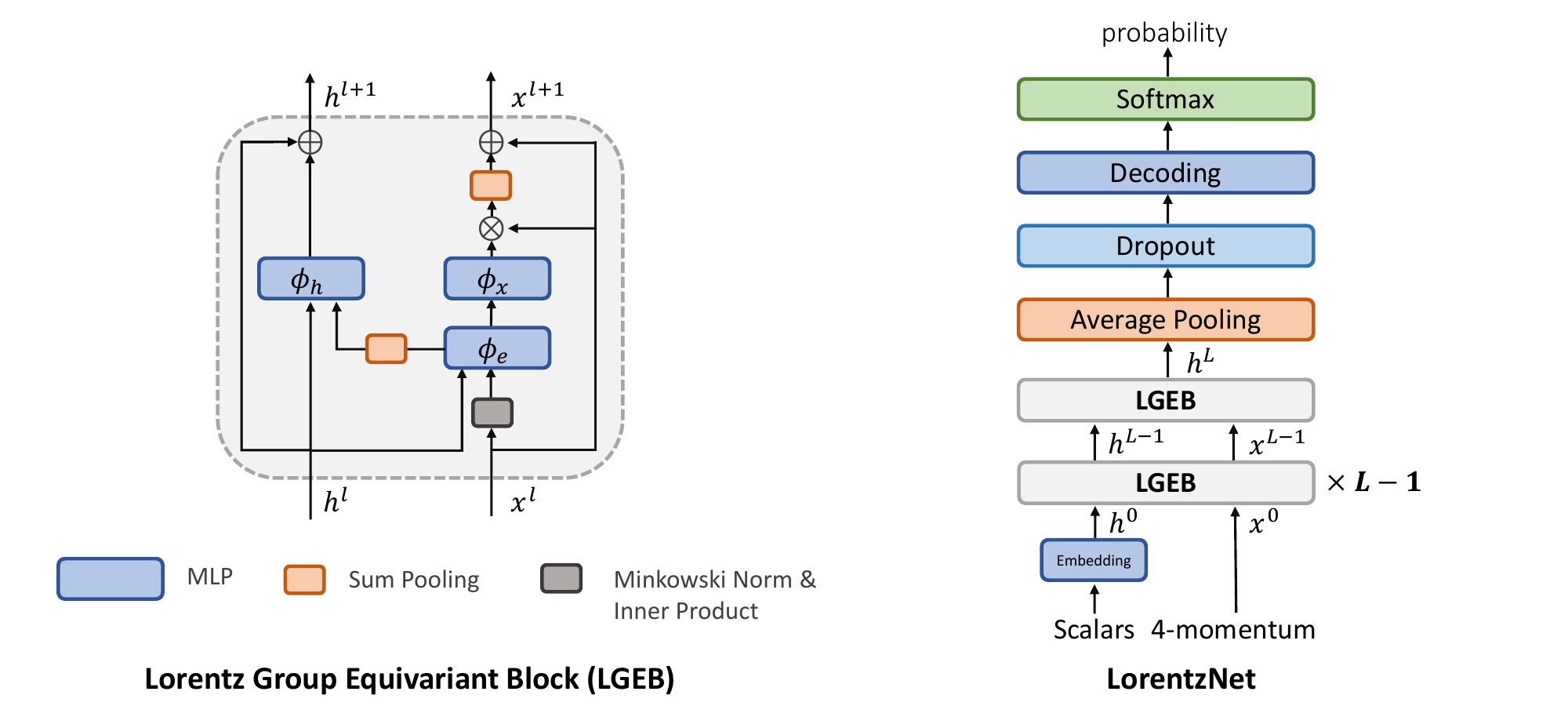}
\caption{The architectures of the Lorentz Group Equivariant Block (left) and LorentzNet (right), as proposed in \citep{Gong:2022lye}.}  \label{fig:LorentzNet}
\end{figure}

An alternative architecture, the LorentzNet, comprising of 6 Lorentz Group Equivalent Blocks, followed by a single fully connected scalar embedding layer, responsible for mapping scalars to a latent space of width 72 is proposed in \citep{Gong:2022lye}. The transformations $\phi_x, \phi_e, \phi_h$ follow the schema: $\mathrm{Linear}(72, 72) \to \mathrm{ReLU} \to \mathrm{BatchNorm1d}(72) \to \mathrm{Linear}(72, 72)$, while $\phi_m$ follows: $\mathrm{Linear(72, 1)} \to \mathrm{Sigmoid}$. The decoding occurs through: $\mathrm{Linear(72, 72)} \to \mathrm{ReLU} \to \mathrm{Linear(72, 2)}$, with a dropout rate of 0.2. The network architecture is articulated in Figure \ref{fig:LorentzNet}. LorentzNet achieves an accuracy of 94.2\% with an AUC of 0.9868. A comparative study of LorentzNet with other representative algorithms shows, only ParticleNet with AUC 0.9858 achieves a similar performance compared to the LorentzNet.
%

A more recent study to generalize the extraction of infrared and collinear safe observables for LHC phenomenology was performed in \cite{Konar:2023ptv} using Hypergraphs. The underlying concept of these hypergraphs are the same as the traditional graphs, but it generalizes the connection between the edges and vertices offering more flexibility in the extraction of event information. 

A variant of geometric data structure, dealing with points represented in some relevant vector space, referred to as point clouds \citep{PntCld1}, has found pertinence in Computer Vision problems, with the advent of algorithms that consume point clouds as inputs and train a DNN architecture over them \citep{PntCld2}. This approach has found its way into object identification in particle collision data using the DeepSet framework, constructing an algorithm accounting for the infrared and collinear safe observables, termed Energy Flow Network (EFN) \citep{Komiske:2018cqr}, which forms a part of the EnergyFlow Python package \citep{Komiske:2017aww}. A similar architectural framework, Particle Flow Network (PFN), from the same group \citep{Komiske:2018cqr}, has been proposed for the same purpose. There, nevertheless, exists a difference between the execution of EFN and PFN, related to the way these formalisms represent and analyze the collision events and the resulting particle interactions within the collider experiments: the EFN architecture primarily focuses on tracking the distribution and transfer of energy resulting from particle collisions with nodes representing specific energy deposits in the calorimeters and the edges depicting the flow of energy between these nodes, indicating the paths taken by particles as they deposit energy in the detector, while the PFN focuses on tracking individual particles produced in the collision from their subsequent interactions with the detector material with the nodes presenting individual particles detected in the collision, (\textit{e.g.},leptons, photons and hadrons) and edges portraying the trajectory taken by these particles through the detector, including interactions with detector material and particle identification.

\begin{figure}[!htb]
\centering
\includegraphics[width=0.4\textwidth]{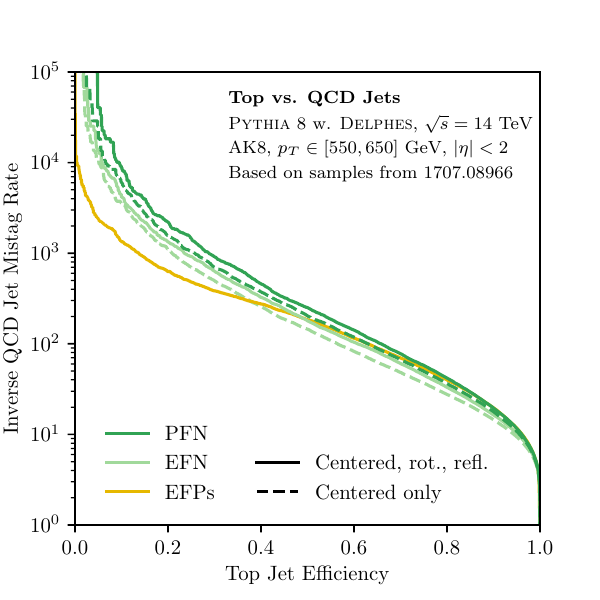}
\caption{The ROC curves comparing the performances of different EFP formalisms introduced in \citep{Komiske:2018cqr}, we refer to the original work for the details on this.}  \label{fig:EFN}
\end{figure}

EFN has been implemented for the segregation of quark-jets from gluon-jets in \citep{Komiske:2018cqr}, using the dataset from \citep{10.21468/SciPostPhys.5.3.028dataEFN} for collisions at a center of mass energy of 14 TeV. Pythia was used for the large-R jet reconstructions ($R = 0.8$) with anti-$k_T$ clustering algorithm, and a jet selection criteria were imposed: $p_{T,jet} ~\in~$[550, 650] GeV, $\lvert \eta \rvert ~<~$2 and $\Delta R_{i,jet} ~<~$0.8, where $i$ runs over all the jet constituents ensuring the existence of all the decay products of top quark inside the jet radius. The jets are then preprocessed following the approach taken in \citep{deOliveira:2015xxd}, and in addition, the net transverse momentum registered in each pixel of the jet-image was normalized. Different approaches associated with graphs, point clouds and computer vision were implemented for the jet image analysis and a comparative study on the performances of different algorithms were presented in Figure \ref{fig:EFN}.
%
%

\subsubsection{Hybrid Top Tagger:}
\begin{figure}[!htb]
\centering   
\includegraphics[width=.3\textwidth]{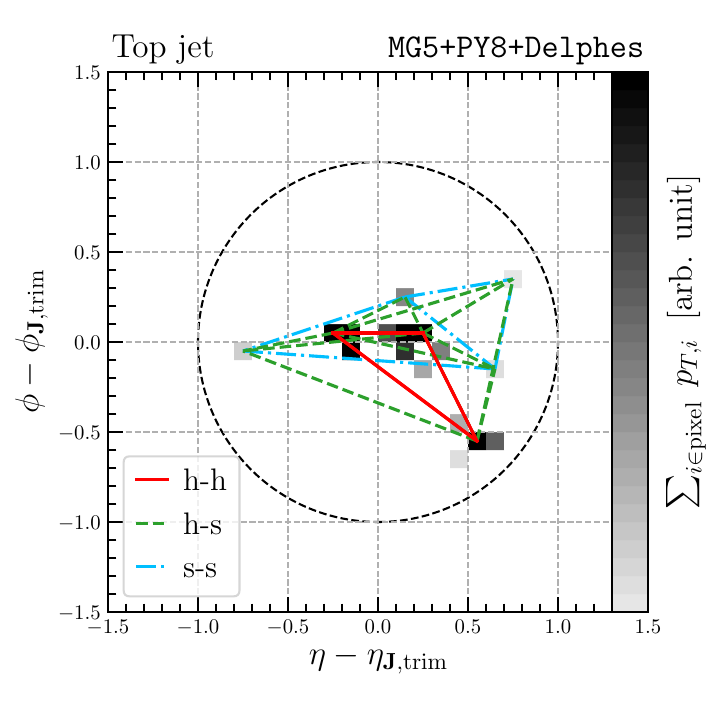}
\includegraphics[width=.3\textwidth]{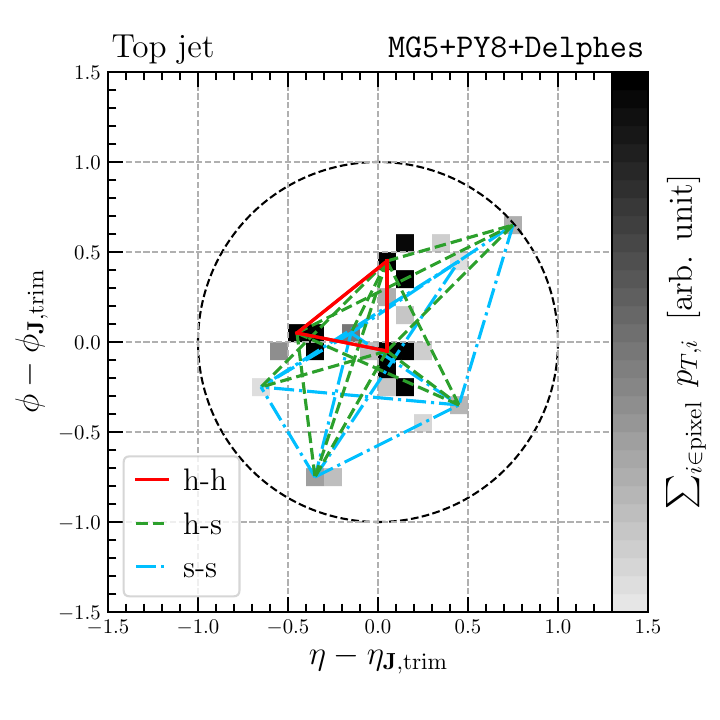} 
\includegraphics[width=.3\textwidth]{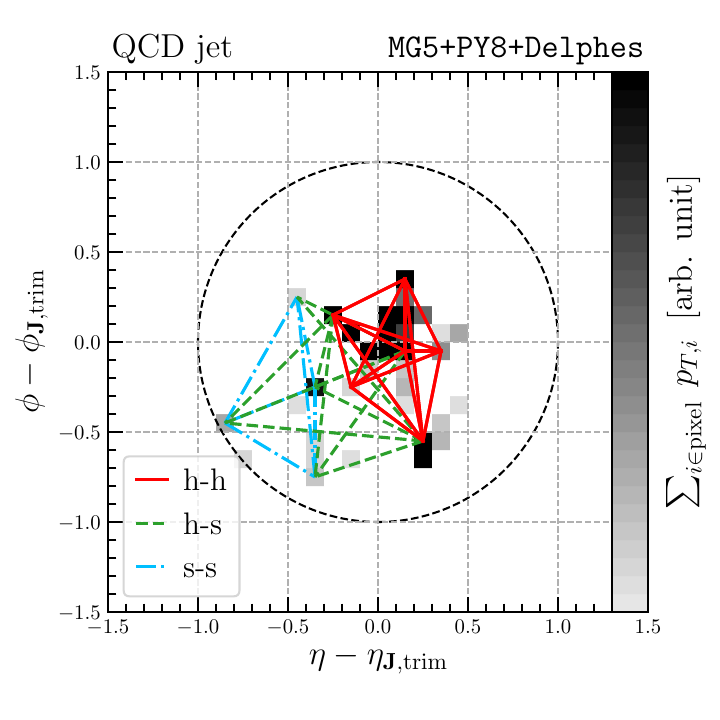} 
\caption{
The graph representations of different jets - the left and middle figures represent the top jets, while the right-most figure for a QCD jet \cite{Chakraborty:2020yfc}. The Lines represent the graphs on the jet images. Different colored lines show the edges between the constituents of the ungroomed and groomed (trimmed) jet.} \label{Fig:jet-image-graph}
\end{figure}

The work done in \citep{Chakraborty:2020yfc} uses a convolution neural network (CNN) for the classification of top against QCD jets($pp\to jj$) using both infra-red and collinear (IRC) safe as well as unsafe observables. IRC safe observables include the two-point energy correlation, jet spectrum as a function of distance between the jet constituents, while IRC unsafe includes Minkowski sequence with the first element of it to be the number of active pixels in a jet image and the second one as the sum of active pixels and adjacent pixels. The top tagger proposed in \citep{Chakraborty:2020yfc} considers pair production of top quark from pp collision ($pp\to t\bar{t}$) in 13 TeV center of mass energy, where the top further decays to $bW$. Madgraph 5 2.6.6 is used for the simulation of parton level events, which were further passed through Pythia 8.226 and Herwig 7.1.3 for parton shower and hadronisation.  Delphes 3.4.1, with the default ATLAS card was employed to consider detector limitations. Reconstructions of the jets were made by fastjet 3.3.0 with anti-$k_T$ clustering algorithm for the radius parameter $R = 1.0$. The tagger was optimised for tagging top-jets satisfying the $p_{T,J} ~\in~ [500,600]$GeV and the mass, $\rm m_J ~\in~ [150,200]$GeV.
This work equally explores the prospect of implementation of graph network architecture, namely, relation network (RN), useful for making two-point correlations, for $1 \to 2$ splitting of partons in the kernel of the parton shower model. A comparative study of the performances between CNN and RN models are performed, and the analysis indicates that the classification ability of RN agrees with that of CNN, upon including IRC unsafe observables, which leads to the inference that the CNN utilises geometric information of soft radiation for the classification task. The RN model is found to outperform its CNN counterpart in terms of stability, and is a more acceptable one for having less complexity in the loss function. The approach taken in \citep{Chakraborty:2020yfc} is intended to implement physics-motivated high-level features in a simple network instead of using low-level features in complex networks. Figure \ref{Fig:jet-image-graph} shows the graph representation of the top and QCD jet images in terms of their respective constituents. A line joining two points, individually articulating the energy deposition of a given jet constituent, represent the graphs on the jet images. The red solid lines represent edges between the constituents of the trimmed jet, while the green dashed lines portray the edges between the constituents of the trimmed jet and the constituents of the cluster not encapsulated in the trimmed jet. The blue dot-dashed lines, in turn, present edges between the pair of constituents of the large-R jet not encapsulated in the trimmed jet.

A recent study based on the two-point energy correlations, Minkowski functionals using jet constituents, and a recursive neural network analyzing subjet constituents was performed in the context of top tagging \citep{Furuichi:2023vdx}. The study observes performance comparable to the Particle Transformer, and, most importantly, the training of the model is reasonably much faster than the transformer network. In addition, the proposed framework also provides some interpretability of the trained model. {Another study worth mentioning is \cite{Sahu:2023uwb}, where the authors have analyzed the importance of leveraging the combined information from the calorimeter towers and tracker at the LHC for ML-based Top Taggers. An extensive examination was conducted on the performance of three machine learning algorithms: the high-level feature (HLF)-based Boosted Decision Tree (BDT), low-level feature (LLF)-based Convolutional Neural Network (CNN), and Graph Neural Network (GNN). The study finds that the information of the distribution and composition of charged and neutral constituents of the jets enhance the performance of the classifiers significantly over a wider $p_T$ range. This enhancement comes due to the different radiation profile of quarks and gluons following QCD principles. However note that this improvement introduces large systematic uncertainties, primarily originating from the different modeling of the parton shower and hadronization, which can be as large as 30-40\%. Interestingly, composite classifiers combining the information of low-level and high-level objects together help in both improving the performance of the classifier and reduce the systematic uncertainty to below 20\%. Additionally, the sensitivity of transverse momentum and mass of the reconstructed jet on the trained models are also examined. 
}

\begin{figure}[!htb]
\centering   
\includegraphics[width=.6\textwidth]{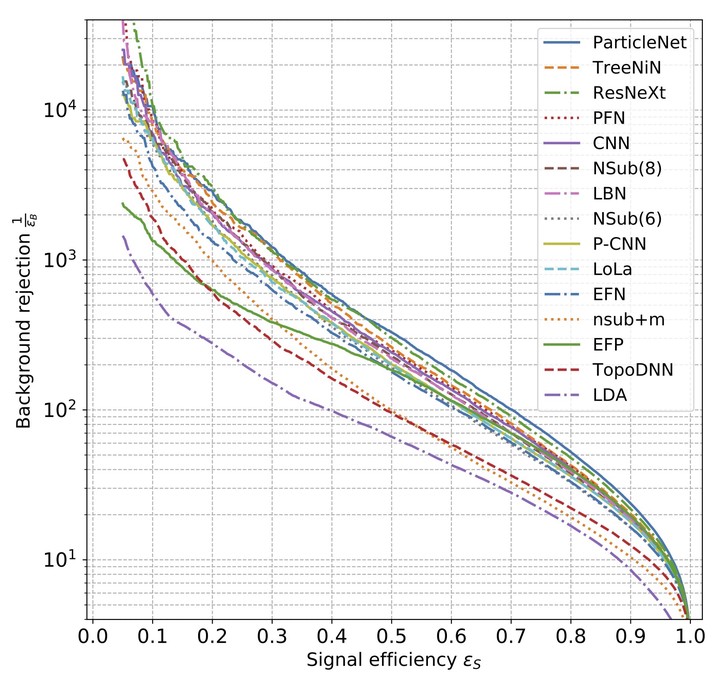} 
\caption{The figure showing the performance of different top taggers using overlaid ROC lines \cite{Kasieczka:2019dbj}.} \label{Fig:result_roc}
\end{figure}

Before we conclude this section, we would like to highlight the results of a comparative study involving  different top-taggers using the same test sample as inputs \citep{Kasieczka:2019dbj}. The ROC curve for different top-taggers are shown in Figure \ref{Fig:result_roc}, estimating the background rejection over signal efficiencies for each of them. As is evident from the figure, the performance of ParticleNet, the 4-vector based tagger, is the best among the rest, with an AUC value of 0.98. In addition, RestNeXt (image-based), TreeNiN (4-vector-based), and PFN (theory-inspired) taggers also exhibit comparable performance. However, we should keep in mind many of the recent-most studies which we discussed previously obtained performances that are comparable (or sometimes better) in background rejection rate.  

{
\subsection{Application of Higgs/Top Taggers in BSM Physics searches}
Having discussed different implementations of Higgs and Top taggers, it is interesting to discuss their application to BSM physics searches. For example, in the context of Type-II Higgs doublet model and Supersymmetric models, the charged Higgs bosons after being produced at the LHC, can decay into $t\bar{b}$ final state. Applying the boosted top jet tagging using jet substructure/ML algorithms, we can first reconstruct the top jet and then reconstruct the charged Higgs mass \cite{Yang:2011jk,Pedersen:2016kyw,Guchait:2018nkp}. 
In theories with extended Higgs sector, boosted SM Higgs bosons can stem from the chain decay of a heavy Higgs state that further decays into bottom-antibottom quark pairs. The proposed BDRS tagger can successfully reconstruct the mass of the SM Higgs boson and thereby the mass of the heavy Higgs boson  
\cite{Chakraborty:2023hrk}. Even boosted boson tagging can help in probing the compressed SUSY scenario with heavier top/bottom squarks \cite{Kang:2017rfw}. On the other hand, Top tagging can be useful for the searches of third generation squarks, namely stop and sbottom, at the LHC; a partial list of studies include  \cite{Goncalves:2016nil,Plehn:2010st,Chakraborty:2013moa,Bhattacherjee:2013tha,Bardhan:2016gui,Bhaskar:2020gkk,Bai:2016zou}. In \cite{Banerjee:2018bio}, the authors use boosted Higgs tagging algorithm to probe Electroweak Precision Physics via boosted Higgs-strahlung at the LHC. Several studies have been performed by the ATLAS and CMS collaborations at the LHC using the boosted Higgs/Top tagging method to probe the new exotic light/heavy particles associated to different BSM physics scenarios; we refer some of the interesting recent-most studies performed using LHC data \cite{
CMS:2017iqt,CMS:2022pjv,CMS:2022sfi,CMS:2022spe,CMS:2022suh,CMS:2022qww,CMS:2022tdo,ATLAS:2020lks,ATLAS:2023taw,ATLAS:2023azi,ATLAS:2023ixh}. Note that, while jet substructure was not initially planned for the LHC experiments, it has become indispensable for the current physics program. The interplay of jet substructure and Machine Learning algorithms is poised to play a crucial role in the dynamic interplay between theoretical advancements and experimental research and development programs in the upcoming colliders \cite{Larkoski:2017jix,Nachman:2022emq}. 
}

\section{Interpretability of the Trained Models}
\label{sec:interpretable}

Machine learning models although highly effective for tasks like classification, regression, clustering, anomaly detection, etc., often pose challenges in terms of complexity.  In the context of HEP, where complex physical processes and particle interactions are studied, interpretability is crucial for gaining insights into the underlying physics and ensuring the reliability of machine learning models. It helps understand the importance of each feature of interest on the outcome, globally, i.e., analyzing the features concerning the whole dataset, as well as locally, i.e.,  understanding the reason for each decision event-by-event. Additionally, there are some physics-informed models, which help in understanding how the model incorporates the physical laws into making decisions, while some also include the ability to estimate the uncertainty in the model prediction.

The performance of deep neural networks, which are one of the most popular choices for object classification purposes, are often achieved at the cost of interpretability. In \cite{Chakraborty:2019imr}, the authors propose an interpretable network trained on the `jet spectrum', denoted by $S2(R)$, which is a essentially an Infra-red and collinear safe two-point correlation function of the jet constituents. An interpretable framework can be obtained by truncating the functional Taylor series constructed using the energy flows. The spectrum as a function of a variable angular scale R reveals the important values (or ranges) of R responsible for classifying Higgs jets from quark/gluon originated jets. The performance of the proposed architecture is not comparable with that of a CNN architecture, but the training time is also significantly less compared to more complex networks like a CNN.  

Lloyd Shapley introduced a method in the 1950s within Cooperative Game Theory, which has been widely applied in enhancing the interpretability of Machine Learning models \cite{2022arXiv220205594R, ShapBook}. This technique, named after him, assigns a value to each player in the coalition, representing their marginal contribution to the coalition's overall worth. The Shapley value is based on the idea of considering all possible permutations of player orderings and calculating the average marginal contribution of each player over all possible coalitions. It takes into account the various ways in which players can join or leave the coalition and how their inclusion affects the coalition's worth. It provides a way to fairly distribute the surplus generated by cooperation among players who might have different levels of contribution or bargaining power.

In the context of particle phenomenology and Machine Learning (ML), the Shapley value contributes to the interpretability of ML models by providing insights into how features (or variables) contribute to the model's predictions \citep{Bhattacherjee:2022gjq, Chowdhury:2023jof, Khot:2022aky}. Shapley values provide an estimate of the importance of different features in the predictions of the ML model. Upon calculating the Shapley value for each feature, one identifies the features with the most significant impact on the model's output. Besides, Shapley values provide a clear comprehension of how individual features interact with each other in influencing the model's predictions. This understanding provides the identification of complex relationships and interactions between different variables in segregation of particular signatures. For event datasets containing numerous features, Shapley values help to identify the most relevant features for making predictions, aiding in feature selection and dimensionality reduction, improving model efficiency and interpretability. Understanding the contribution of each feature to the model's predictions assists in validating the model's decisions, ensuring its reliability and enhancing the interpretability, transparency, and trust in the models' predictions, ultimately advancing our understanding of fundamental particles and their interactions. In Ref.\cite{Bhattacherjee:2022gjq}, the authors study the top taggers using N-Subjettiness ratios and several observables of the Energy Correlation functions as input features to train the eXtreme Gradient BOOSTed decision tree (XGBOOST). The effect of matching the parton level top and its decay products to the reconstructed boosted top jet are found to be significant. Using Shapley values, it is found that along with individual contributions, the interaction effects (i.e., cross correlations) also control the performance of the tagger. This in turn motivates the author to propose a hybrid top tagger combining the important input features. 

Even though for top tagging, majority of the analysis focus on the case when the boosted top quark either decays hadronically or leptonically, in certain BSM theories, a top quark can decay to a charm quark and a Higgs boson which eventually decays to a pair of b-quarks. This decay is extremely rare in SM, as flavor changing neutral currents are highly suppressed at the tree level. However, even though branching ratios are low, no existing top tagger would be sensitive to probe such decay modes of top quark at the boosted regime. The authors in Ref.\cite{Chowdhury:2023jof} made an attempt to build a top tagger which specifically focuses on this particular decay mode. A MultiVariate Analysis using two different boosting algorithms, viz., XGBoost and AdaBoost, the authors perform a comparative analysis on the performance of the proposed top tagger for the two boosting algorithms along with conventional cut-based top tagging algorithms. The Shapley values are also calculated to articulate the contribution of observables towards appropriate identification of the signal and background events, and consequently providing an insight of the trained model.

Layerwise relevance propagation (LRP) method is one of most popular methods of interpreting the decision of machine learning models based on neural networks. A framework was introduced to extract and understand the decision-making information of top tagger build from a deep neural network (DNN) architecture using substructure based observables \cite{Agarwal:2020fpt}. The authors aimed to identify the expert variables that augment inputs (``eXpert AUGmented" variables, or XAUG variables), and then apply LRP method to networks both with and without XAUG variables. The LRP technique finds the relevant information the network is using, and helps in improving the network performance significantly when XAUG variables are introduced to the network. A similar analysis using LRP method in the context of a 
GNN based particle-flow (PF) algorithm has been performed to identify the relevant nodes and features for its predictions \cite{Mokhtar:2021bkf}.

One of the most comprehensive studies in the context of top tagging models was performed in \cite{Khot:2022aky}. There exists several methods of explainable AI (XAI) which helps in exploring the working principles of different deep neural networks, especially it provides the all important input-output relationship present in the data. Considering a few important mathematical frameworks of XAI, namely 
the $\Delta$AUC method, Shapely Additive Explanations (SHAP), Layerwise Relevance Propagation (LRP), and Neural Activation Pattern (NAP) Diagrams, the authors have showed a number of important aspects connecting the data with the trained models. In fact, the authors have cautioned on making an interpretation from the feature importances/ranking without making a thorough study especially for the cases when the input features are highly correlated. Each of those frameworks has some advantages over the other choices. For example, while the RNA scores provide better understanding of how information is transferred through different layers of the network, 
the NAP diagrams help in estimating the uncertainties in event classification. The study of these interpretable networks also leads to propose a hybrid interpretable network, called the Particle Flow Interaction Network (FPIN) which not only outperforms (or, at least comparable to) the existing top tagging models, but the network also consists of a smaller number of parameters leading to quicker training and better convergence rate. It is noteworthy that a similar strategy for better and faster training with less number of trainable parameters has been proposed based on Distance Correlation (DisCo) \cite{Das:2022cjl}.

A versatile network, called permutation equivariant and Lorentz invariant or covariant aggregator network, in short PELICAN, that takes into account symmetry principles (specifically full Lotentz symmetry) was proposed to reduce the complexity of the architecture and increase the interpretability without compromising the performance \citep{Bogatskiy:2023nnw}. The PELICAN network outperforms many traditional taggers especially designed for tagging boosted top quarks and quarks/gluons with fewer number of trainable parameters. In fact, the network achieves similar classification performance is 5 different categories of jets (gluon jets, light quark jets, $W$-boson, $Z$-boson, and top quark jets) showing efficient utilization of training parameters. Furthermore, the network is capable of reconstructing charge particle track, W-boson mass from top quark decay and even reconstructing the full collision event. The combination of the full Lorentz symmetry and permutation symmetry helps in explaining the PELICAN weights as the clustering coefficients which in turn makes the network explainable.

The Shower Deconstruction methodology plays the central role in distinguishing between signal and background jets by leveraging the details of parton shower. This method, however, suffers from combinatorics issues as the number of jet constituents increases. In Ref.\cite{Ngairangbam:2023cps}, the authors propose a novel method (likelihood analysis) of identifying the most probable shower history as a Markov Decision Process (MDP). Employing a sophisticated modular point-transformer architecture, the proposed algorithm efficiently learns the optimal policy and the resulting neural agent excels in constructing likely shower histories with robust generalization capabilities on unseen test data. More importantly, this method reduces the inherent complexity of the inference process, and thereby establishes a linear scaling relationship with the number of constituents of a given jet.

As we have already discussed in previous sections, the majority of the studies providing remarkable improvements using machine learning algorithms over the traditional methods analyse low-level features. However, this methodology often lacks interpretability, therefore interpretable architectures with transparent decision making processes are desperately needed. One such attempt was made very recently in Ref.\cite{Furuichi:2023vdx}, where authors have proposed an Analysis Model (AM) that processes a few selected high level features designed to capture important features of the top jets. The model is set up in a modular structure which includes a relation network that examines two-point energy correlations, mathematical morphology, and Minkowski functionals, and a recursive neural network involving subjet multiplicity and their color charges. The proposed framework obtains performance comparable to the Particle Transformer (ParT) in top jet tagging with smaller training uncertainties.

\section{Summary and Outlook}
\label{sec:summary} 

Through the years of experimental and theoretical studies, the SM of particle physics, a theory which unifies the fundamental constituents of Nature and their interactions, found to be mostly successful in explaining the Nature except in a few cases. In this quest, several BSM studies have been performed, where the BSM events are searched over the SM-driven events. Jets and their substructure analysis in the boosted regime are proven to be providing remarkable success in this thrive. With the increase of luminosity in the LHC, the study of SM has also found new directions, especially in the boosted regime. Though the cut-based techniques left a prominent footprint in the SM and BSM searches, the vigorous development of ML and AI, added new heights to it. This work summarizes the development of different tagging algorithms in the ML era, especially in the light of top and Higgs tagging. 

One of the most initial and effective ML algorithms in the area of High energy physics was the Decision tree algorithm. Similar to the cut-based techniques, it splits the phase space based on the features in an event iteratively thereby constructing a tree-like structure containing a set of events in each final leaf node. The sets are representations of different classes, differentiating BSM events of interest from similar irreducible backgrounds of SM processes. A further development to the decision tree algorithm is the Boosted Decision Tree, where the misclassified events in the previous iteration of training get more focus in the next training, improving the overall training performance.

The more advanced development towards ML was seen around 1940's, when by mimicking the human brain a new algorithm entered the ML paradigm. In this new algorithm, except for the input and output layer, a hidden layer with a finite number of perceptrons is employed. Similar to brain neurons, these perceptrons are designed such that the network can decode any complex functions hidden within the patterns of the values of the features. These feed-forward networks with one hidden layer are termed as Artificial Neural Networks or ANN. The training process of these networks aim to minimize the error in assigning the weights of each neuron in forward propagation by updating them during backward propagation. 

A further improvement in the result of NN is seen, when implemented with multiple hidden layers, leading to a new network architecture termed as Deep Neural Network or DNN. In the view of jet tagging, especially Higgs and Top tagging, we find that the DNN with more hidden layers with less number of nodes sometimes performs better than the DNNs with a large number of nodes comprised in a smaller number of hidden layers. One step ahead the DNN is further improvised by taking a combination of convolutional layers. These networks termed as CNN are established to be good at understanding complex structures of the boosted jets without any feature engineering. Sliding a kernel or learnable filter throughout the jet image, the CNN architecture captures the pattern in an image. This understanding helps to segregate between the jet image of the process of interest from backgrounds of similar kind.  It is found through several studies that increasing the feature map and the number of nodes of dense layers improve the top tagging efficiency over mistagging rate significantly. Additionally it is also found that this network trained with low level features performs almost 3 times better than the BDT trained with high level features. However, the taggers like DeepTopLoLa which uses Lorentz Vectors as input and adds a Lorentz layer to the DNN for converting the vectors to relevant kinematic variables, can produce similar results.

In this review, we first discuss the traditional tagging method of boosted Higgs jet, namely the BDRS algorithm, and then highlighted the observation of different studies using state-of-the-art machine learning algorithms, starting with constituent based methods, image based algorithms using CNNs, combination of high level objects and image based analysis, Lund Plane, and more recent-most ones using GNN architectures. In the majority of these studies, a significant amount of improvement in the tagging performance has been observed while using these advanced methods. Similar observations were also found while analyzing the top quarks in the boosted regime through top tagging methods. We discussed that even though traditional substructure based algorithms as implemented in HEPTopTagger performed reasonably well in moderate to large $p_T$ regime of the top quark, advanced machine learning involved algorithms based on jet images, constituent based RNN architecture, or graph based ones provide amazing performance in classifying top jets from quark/gluon jets.

These sophisticated algorithms, even though extremely powerful in classification, regression or clustering events, come with a lot of complexity, a large number of trainable free parameters, and most importantly, lack in interpretability. Several attempts have been made in the recent past to propose trainable models that are explainable, so that, along with improved performance at a given task, we also understand the decision making process in achieving the result. Some of these methods that help in explaining a neural network are SHAP analysis, Layerwise Relevance Propagation method, Neural activation pattern etc. We have highlighted several studies involving these explainable algorithms, along with their merits and demerits. Tremendous amounts of efforts are ongoing to extend the applicability of these novel methods, diving deeper into their adaptability and usefulness across different data structures, especially in complex structures like graph nets and transformers. 
These will certainly revolutionize the current landscape of Higgs and Top taggers, and also broaden generic jet classification scenarios.

\section{Acknowledgements}
The work of AC is funded by the Department of
Science and Technology, Government of India, under Grant No. IFA18-PH 224 (INSPIRE Faculty
Award).


\vskip 1.5 cm 

\noindent
{\bf \large Data Availability Statement}: No Data associated in the manuscript.

\newpage

\bibliography{Ref}

\end{document}